\documentclass[11pt,a4paper]{article}
\pdfoutput=1

\usepackage{amsmath, amssymb, graphics, setspace}
\usepackage{jheppub}

\usepackage[T1]{fontenc}
\usepackage[utf8]{inputenc}
\newcommand{\Ultimes}{^^e2^^8b^^89}
\newcommand{\UbbR}{^^e2^^84^^9d}
\newcommand{\Usupeight}{^^e2^^81^^b8}
\newcommand{\Usubtwo}{^^e2^^82^^82}

\usepackage{tensor}

\newcommand {\mathsym}[1]{{}}
\newcommand {\unicode}[1]{{}}

\usepackage{slashed}
\usepackage{youngtab}

\newcommand{\rd}{\ensuremath{\mathrm{d}}}

\newcommand\nn{\nonumber}

\newcommand{\beq}{\begin{equation}}
\newcommand{\eeq}{\end{equation}}

\newcommand{\beqa}{\begin{eqnarray}}
\newcommand{\eeqa}{\end{eqnarray}}

\newcommand{\bn}{\begin{equation}}
\newcommand{\en}{\end{equation}}
\newcommand{\by}{\begin{eqnarray}}
\newcommand{\ey}{\end{eqnarray}}

\newcommand{\Ga}{\Gamma}

\newcommand{\ep}{\epsilon}

\definecolor{RedViolet}{RGB}{167,64,178}

\hypersetup{unicode=true} 
\newcommand{\texorpdfGtwo}{\texorpdfstring{$G_2$}{G\Usubtwo}}
\newcommand{\texorpdfGtwoReight}{%
  \texorpdfstring{$G_2\ltimes \mathbb{R}^8$}%
  {G\Usubtwo\ \Ultimes\ \UbbR\Usupeight}}

\def\SU(#1){\ensuremath{\mathrm{SU}(#1)}}
\def\Spin(#1){\ensuremath{\mathrm{Spin}(#1)}}

\let\latexstar\star
\renewcommand{\star}{\mathbin{}\mathord{\latexstar}}

\title{Supersymmetric geometries of  IIA supergravity III}

\author[a]{Ulf~Gran,}
\author[b]{George Papadopoulos,}
\author[a]{Christian von Schultz}
\affiliation[a]{Department of Physics\\Division for Theoretical Physics\\ Chalmers University of Technology\\
SE-412 96 Göteborg, Sweden}
\affiliation[b]{Department of Mathematics\\
King's College London\\
Strand\\
London WC2R 2LS, UK\\}
\emailAdd{ulf.gran@chalmers.se}
\emailAdd{george.papadopoulos@kcl.ac.uk}
\emailAdd{christian.von.schultz@chalmers.se}

\abstract{ We find that (massive) IIA backgrounds that admit a $G_2\ltimes \mathbb{R}^8$ invariant Killing spinor  must exhibit a null Killing vector field which leaves the Killing spinor invariant and that the rotation of the Killing vector
field satisfies a certain $\mathfrak{g}_2$ instanton condition. This result together with those in \cite{iiaI} and \cite{iiaII} complete the classification of geometries of all (massive) IIA backgrounds that preserve one supersymmetry.
We also explore the geometry of a class of backgrounds which admit a $G_2\ltimes \mathbb{R}^8$ invariant Killing spinor and where in addition  an appropriate
1-form bilinear vanishes.   In all cases, we express the fluxes of the theory in terms of the geometry. }


\begin{document}
\maketitle


\section{Introduction}

It has been known for some time that the maximally supersymmetric backgrounds of standard IIA supergravity are locally isometric to Minkowski space,
while massive IIA supergravity does not admit such solutions \cite{maxsusy}.  Furthermore, it has been shown in \cite{bandos} that there are no IIA solutions
that preserve strictly 31 supersymmetries following a similar proof for IIB in \cite{iibmaxsusy}. On the other end,  there are locally four
different types of IIA backgrounds preserving one supersymmetry distinguished by the isotropy group of their Killing spinor in the local gauge group $\Spin(9,1)$.  The geometry
of the backgrounds with isotropy groups $\Spin(7)$, $\Spin(7)\ltimes \mathbb{R}^8$ and $\SU(4)$ have been investigated in \cite{iiaI, iiaII}.  The purpose of this
paper is to examine the geometry of the remaining backgrounds, which admit a Killing spinor with isotropy group $G_2\ltimes \mathbb{R}^8$. The combined   four cases
give a complete description of the local geometry of all (massive) IIA backgrounds that preserve one supersymmetry.

The methodology we use to solve the Killing spinor equations (KSEs) of IIA supergravity is that of spinorial geometry proposed in \cite{spingeom}.  For this we choose
a representative of the $G_2\ltimes \mathbb{R}^8$-invariant Killing spinor as
\by
\epsilon=f (1+e_{1234})+ g (e_1+e_{234})~,
\label{ksg2}
\ey
where $f$ and $g$ depend on the spacetime coordinates. If either $f$ or $g$ vanishes, then the isotropy group of $\epsilon$ becomes $\Spin(7)\ltimes \mathbb{R}^8$ and
such solutions have already been investigated in \cite{iiaI}. Because of this we take that neither $f$ nor $g$ vanishes on the patch where the KSEs are investigated.
 Then this spinor is substituted into the KSEs, which turn into a linear system for components of the fluxes and the
spin connection, which encodes the geometry. The solution of the linear system expresses some of the fluxes in terms of the geometry and also gives the conditions on the
geometry for a background to admit such a Killing spinor.

In particular, we find that all backgrounds for which (\ref{ksg2}) is  a Killing spinor satisfy the following geometric conditions
\by
{\cal L}_K g=0~,~~~{\cal L}_K \epsilon=0~,~~~\rd e^-\in \Gamma(\Lambda'_{\bf 7}\oplus I\oplus \Lambda_{\bf 14})~,~~~g(K,K)=0~.
\label{geomg2}
\ey
Therefore, such backgrounds admit a null 1-form spinor bilinear $e^-$ such that the associated vector field $K$ is Killing  and leaves invariant the Killing
spinor
$\epsilon$. These two conditions have been expected as they follow from the properties of  the symmetry superalgebra for supersymmetric backgrounds\footnote{To our knowledge the  investigation
of the symmetry superalgebra for massive IIA backgrounds has not been carried out in the generality presented for IIB backgrounds in \cite{jose}.}, see e.g.~\cite{jose}. In addition, the rotation $\rd e^-$ of $K$ is restricted to be a section of a bundle $\Lambda'_{\bf 7}\oplus I\oplus \Lambda_{\bf 14}$ which will be defined in section \ref{secgeo},
  and which can be interpreted as a $\mathfrak{g}_2$ instanton condition with an additional null component.
 Furthermore, the geometry of spacetime
satisfies two more mild geometric conditions given by (\ref{fgconxx}) and (\ref{fgconbxx}). The latter can also be interpreted as expressing
 the dilaton $\Phi$ in terms of the geometry.

There is a special subclass of $G_2\ltimes \mathbb{R}^8$ backgrounds which is characterized by the additional condition that a certain
1-form bilinear vanishes.  It turns out that the IIA backgrounds admit two 1-form bilinears $\kappa(\epsilon, \epsilon)$ and $\kappa(\epsilon, \Gamma_{11}\epsilon)$,
and that
\by
\kappa(\epsilon, \Gamma_{11}\epsilon)= h\, \kappa(\epsilon, \epsilon)~,
\ey
where $h$ is a spacetime function and $\kappa(\epsilon, \epsilon)$ is the bilinear associated with the Killing vector field $K$. The requirement that $\kappa(\epsilon, \Gamma_{11}\epsilon)$
vanishes leads to the condition that $h=0$ which in the gauge (\ref{ksg2})  implies that $f^2=g^2$.  In this special case, the conditions on the geometry are as in
(\ref{geomg2}) and
\by
(\star\varphi, \rd\varphi)=0~,
\label{geombg2}
\ey
where $\varphi$ is the fundamental $G_2$ 3-form and $\star\varphi$ is its dual as defined in appendix~\ref{sforms}\@. This vanishing condition
is well-known as it implies that one of the classes in a Gray-Hervella type of classification of manifolds with $G_2$ structure vanishes \cite{stefan}.
There is an additional geometric condition given in (\ref{fgconbxxx}), which is the analogue of (\ref{fgconbxx}) for the special case.  Again this can also be interpreted as a condition on the dilaton. Furthermore,
 the expression of the fluxes in terms of the geometry dramatically simplifies.

The description of the geometry we have given is local, i.e.~it assumes that the Killing spinor can always  be brought into the form (\ref{ksg2}) up to
a local Lorentz transformation. Generically, solutions of the KSEs are not required to preserve  the isotropy group of the Killing spinors in $\Spin(9,1)$ everywhere on the spacetime.
 As a result
there may be backgrounds which (globally) admit Killing spinors whose isotropy groups differ between different patches. Furthermore,
as the rank of the IIA spin bundle is 32, and thus much larger than the dimension of the spacetime, the mere existence of a non-vanishing spinor
does not reduce the structure group of the spacetime\footnote{From this perspective, G-structures are not an effective tool
to globally describe the solutions of the KSEs for type II backgrounds.}.  Of course if one asserts that the isotropy group of the Killing spinor is the same everywhere
 on the spacetime, then the structure group reduces to a subgroup of the isotropy group. However, this is an additional assumption.

This paper has been organized as follows. In section \ref{secsolkse}, we summarize the KSEs of (massive) IIA supergravity and describe the solution of the linear system,
for all backgrounds admitting a $G_2\ltimes \mathbb{R}^8$-invariant Killing spinor, for both the fluxes and geometry.  In section \ref{secspecial}, we investigate the geometry
of the special class of backgrounds for which the aforementioned 1-form bilinear vanishes. In section \ref{secconclusion}, we give our conclusions.  In appendices~\ref{conv} and~\ref{sforms}, we describe various
formulae which are useful for the analysis that follows and we give explicitly the expressions for all the form bilinears of the spinor (\ref{ksg2}). In appendix~\ref{g2sol}, we give the linear system organized
in $G_2$ representations.  In appendices~\ref{linsyssol} and~\ref{appspec}, we present the solution of the linear system  for generic as well as for special $G_2\ltimes \mathbb{R}^8$ backgrounds, respectively.


\section{Solution of the Killing spinor equations}
\label{secsolkse}

\subsection{Killing spinor equations}

The KSEs of (massive) IIA supergravity are the vanishing conditions of the supersymmetry variations of the fermions of the theory \cite{Giani:1984wc,Campbell:1984zc, Huq:1983im, romans,Bergshoeff:2010mv}  evaluated at the locus where all fermions vanish. The conventions for the fields that we shall use, including the field equations
 and Bianchi identities, are as those in \cite{iiaI} which closely follow those of \cite{Bergshoeff:2010mv}. For completeness we state the KSEs of the theory, which
  are the vanishing conditions of
\by\label{KSE}
{\cal D}_M \ep &\equiv& \nabla_M \ep + \tfrac{1}{8} H_{MP_1 P_2}\Ga^{P_1 P_2}\Ga_{11}\ep +\tfrac{1}{8}  S\Ga_M \ep \notag \\
&& +\tfrac{1}{16} F_{P_1 P_2}\Ga^{P_1 P_2}\Ga_M \Ga_{11} \ep +\tfrac{1}{8\cdot 4!} G_{P_1 \cdots P_4}\Ga^{P_1 \cdots P_4}\Ga_M \ep ~,\nn\\
{\cal A} \epsilon &\equiv & \partial_P \Phi \Gamma^P \ep + \tfrac{1}{12} H_{P_1 P_2 P_3}\Ga^{P_1 P_2 P_3}\Ga_{11}\ep +\tfrac{5}{4}  S \ep \notag \\
&& +\tfrac{3}{8} F_{P_1 P_2}\Ga^{P_1 P_2}\Ga_{11}\ep +\tfrac{1}{4\cdot 4!} G_{P_1 \cdots P_4}\Ga^{P_1 \cdots P_4} \ep ~,
\ey
where $\nabla$ is the spin connection, $H$ is the NS-NS 3-form field strength, $S, F,  G$ are the RR $k$-form field strengths, $k=0, 2 ,4$,  and $\Phi$ is the dilaton.
Note that the RR form field strengths in the KSEs have been rescaled with the exponential of the dilaton.

The spinor $\epsilon$ is in the Majorana representation of $\Spin(9,1)$, and $\epsilon$ is from now on taken to be commuting. The first KSE in  (\ref{KSE}) arises from the supersymmetry variation of the gravitino of the
theory and is a parallel transport equation,  while the second condition  arises from the supersymmetry
variation of the dilatino and is an algebraic equation on $\epsilon$.  In what follows, we shall seek solutions
to the conditions ${\cal D}\epsilon={\cal A}\epsilon=0$ without making simplifying assumptions regarding the fields, the Killing spinor or the geometry of spacetime.
For spinors,  we use the same conventions as those employed in type IIB supergravity in the context of spinorial geometry \cite{iib1}, e.g.~we choose
$\Gamma_{11} = - \Gamma_{0 1 \cdots 9}$.


\subsection{Solution of the KSEs for the \texorpdfGtwoReight{} backgrounds}

\subsubsection{Linear system and geometry}

Applying spinorial geometry to the KSEs of (massive) IIA supergravity (\ref{KSE}) for the spinor (\ref{ksg2}) turns these equations into a linear system with
variables given by the components of the fluxes. This linear system is originally expressed in $\SU(3)$ representations and it is rather involved. This is because
the expression for the Killing spinor (\ref{ksg2}) is manifestly $\SU(3)$ invariant.  However, after
some extensive computation the linear system can be recast into $G_2$ representations and this is given in appendix~\ref{g2sol}\@. The equations of the linear system depend on the fundamental $G_2$ forms
$\varphi$ and $\star \varphi$. As will become apparent below this is required   so that, after a decomposition in $G_2$ representations, the equations of the linear system depend on the appropriate components of  the fluxes and geometry.

To elucidate the notation and also give some preliminary description of the geometry of spacetime, note that  all supersymmetric backgrounds
which admit a Killing spinor with isotropy group $H\ltimes \mathbb{R}^m$, where $H$ is a compact group,   admit a no-where vanishing null 1-form $e^-$ for which its associated
vector field $K=g^{-1} e^-$ is Killing. Such a 1-form splits the tangent space of the spacetime $M$ as
\by
0\rightarrow \mathrm {Ker}\, e^-\rightarrow TM\xrightarrow{e^-} I\rightarrow 0~,
\ey
where $I$ is the trivial real line bundle.
To understand $\mathrm {Ker}\, e^-$, we choose a vector $W$ such that $e^-(W)=1$ and define the 1-form $e^+(Y)=g(W,Y)$.  Clearly $e^+$ is not uniquely defined
but all considerations are independent of the choice of $e^+$.  The metric $g$ in such a case can be decomposed as
\by
\rd s^2= 2 e^- e^++ \rd s^2_\perp
\ey
where $\rd s^2_\perp(K,Y)=\rd s^2_\perp(W,Y)=0$.  Then $\mathrm {Ker}\, e^-$ is spanned by $K$ and all vectors $Z$ which are perpendicular to both $K$ and $W$, i.e.~$g(Z,K)=g(Z,W)=0$.
To decompose the flux form field strengths, one considers
\by
0\rightarrow \mathrm {Ker}\, K\rightarrow T^*M\xrightarrow{K} I\rightarrow 0~.
\ey
In such a case $\mathrm {Ker}\, K$ is spanned by $e^-$ and the 1-forms $e^I$ which are orthogonal to both $e^-$ and $e^+$.  The metric can then be written as
\by
\rd s^2= 2 e^- e^++ \delta_{IJ} e^I e^J~.
\ey
Furthermore, a $k$-form flux can be decomposed as
\by
{\mathcal F}^k=\frac{1}{2}e^+\wedge e^-\wedge{\mathcal F}^{(k-2)}_{+-} +e^+\wedge {\mathcal F}^{(k-1)}_{+} +e^-\wedge {\mathcal F}^{(k-1)}_{-} +{\mathcal F}^{(k)}~,
\ey
where $(k)$, $(k-1)$ and $(k-2)$ denote the degree of the form in the directions orthogonal to both $e^+$ and $e^-$.

In the $G_2\ltimes\mathbb{R}^8$ case that we are investigating, there is an additional distinct 1-form $e^1$ that arises.
This is due to the choice of the representative (\ref{ksg2}) of the Killing spinor. On the other hand,
if one assumes that the spacetime globally admits a $G_2\ltimes\mathbb{R}^8$ invariant Killing spinor, the structure group of the spacetime reduces to a subgroup
of $G_2$ so topologically the tangent space splits as $TM=I^3\oplus E$.  There are thus three no-where vanishing vector fields on the spacetime
which can be chosen such that their associated 1-forms are $e^+, e^-, e^1$.  In either case, the metric and $k$-form fluxes
can be decomposed as
\by
\rd s^2=2 e^+ e^-+ (e^1)^2+ \rd s^2_{(7)}~,~~~\rd s^2_{(7)}=\delta_{ij} e^i e^j~,
\label{g2metr}
\ey
where $e^i$ span the seven directions orthogonal to $e^+, e^-, e^1$ in $E$,
and
\by
{\mathcal F}^k=\frac{1}{6}  e^a\wedge e^b\wedge e^c\wedge {\mathcal F}^{(k-3)}_{abc}+\frac{1}{2}e^a\wedge e^b\wedge {\mathcal F}^{(k-2)}_{ab}  +
 e^a\wedge {\mathcal F}^{(k-1)}_{a}+ {\mathcal F}^{(k)}~,
\label{fdec}
\ey
where $a,b,c=+,-,1$ and the superscript in brackets $(n)$ denotes the degree of the form along $E$. The components of the fluxes, as well as those of the spin connection
$\Omega$, that appear in the linear system in appendix~\ref{sforms} are defined according to the above decomposition.

\subsubsection{Solution for the fluxes}

The components of the fluxes along $E$ can be further decomposed in $G_2$ representations. As $G_2\subset \Spin(7)$ acts with the fundamental 7-dimensional
representation on the typical fibre of $E$, the 2-forms decompose as ${\bf 7}\oplus {\bf 14}$, and the 3- and 4-forms decompose as ${\bf 1}\oplus {\bf 7}\oplus {\bf 27}$.
 Using these decompositions, the 4-form, 3-form and 2-form fluxes can be written as
 \by
 G&=&\frac{1}{6} e^a\wedge e^b\wedge e^c\wedge G^{\bf 7}_{abc} +\frac{1}{2} e^a\wedge e^b\wedge(G^{\bf 7}_{ab} + G^{\bf 14}_{ab})+ e^a\wedge(G^{\bf 1}_a+ G^{\bf 7}_r+G^{\bf 27}_a)
 \cr &&~~~~~~~+
 G^{\bf 1}+ G^{\bf 7}+G^{\bf 27}~,
 \ey
 \by
 H=\frac{1}{6} H^{\bf 1}_{abc} e^a\wedge e^b\wedge e^c+\frac{1}{2} e^a\wedge e^b\wedge H^{\bf 7}_{ab} +e^a\wedge (H^{\bf 7}_a+ H^{\bf 14}_a) +
 H^{\bf 1}+ H^{\bf 7}+H^{\bf 27}~,
 \ey
 and
 \by
 F=\frac{1}{2}F^{\bf 1}_{ab}  e^a\wedge e^b+ e^a\wedge F^{\bf 7}_a +
 F^{\bf 7}+ F^{\bf 14}~,
 \ey
 respectively, where we have suppressed the degree of the forms along $E$.

The solution of the linear system as described in appendix~\ref{linsyssol} expresses some of the above components of the fluxes in terms of geometry as well as
in terms of some other components of the fluxes which are not restricted by the KSEs.
To relate the above decomposition of the fluxes in $G_2$ representations with the expressions in appendix~\ref{linsyssol} observe that the decomposition of 2-forms, 3-forms and 4-forms along $E$
can be written as follows.
For the 2-forms, we have
\begin{eqnarray}
\chi^{(2)}=\frac{1}{2} \chi^{\bf 7}_i \varphi^i{}_{jk} e^j\wedge e^k+ \chi^{\bf 14}~,
\end{eqnarray}
where $\chi^{\bf 14}_{ij} \varphi^{ij}{}_k=0$ and
\by
\chi^{\bf 7}_i=\frac{1}{6} \chi^{(2)}_{mn} \varphi_i{}^{mn}~.
\ey
For the 3-forms, we obtain
\by
\chi^{(3)}&=&\chi^{\bf 1} \varphi+\frac{1}{6} \chi^{\bf 7}_i \star\varphi^i{}_{jkl} e^j\wedge e^k\wedge e^l+ \frac{1}{6}  \chi^{\bf 27}_{i[j} \varphi^i{}_{kl]}e^j\wedge e^k\wedge e^l~,
\ey
where
\by
&&\chi^{\bf 1}=\frac{1}{42} \chi^{(3)}_{ijk} \varphi^{ijk}~,~~~\chi^{\bf 7}_i=\frac{1}{24} \chi^{(3)}_{kmn} \star\varphi_i{}^{kmn}~,~~~
\cr
&&
\chi^{\bf 27}_{ij}=\frac{3}{4} \big(\chi^{(3)}_{mn(i}\varphi_{j)}{}^{mn}- \frac{1}{7} \delta_{ij}\chi^{(3)}_{kmn} \varphi^{kmn}\big)=\frac{3}{4}\chi^{(3)}_{mn(i}\varphi^{\phantom(}_{j)_o}{}^{mn}~,
\ey
and the notation $(\cdot\cdot)_o$, which we adopt from now on, denotes the symmetric  traceless part of the two indices in the round brackets.

Similarly for the 4-forms, we have
\by
\chi^{(4)}&=&\chi^{\bf 1}\star\varphi+\frac{1}{4!} \chi^{\bf 7}_{i_1} \varphi_{i_2i_3i_4} e^{i_1}\wedge \dots \wedge e^{i_4}+\frac{1}{4!} \chi^{\bf 27}_{\ell i_1}
\star\varphi^\ell{}_{i_2i_3 i_4} e^{i_1}\wedge \dots \wedge e^{i_4}~,
\ey
where
\by
&&\chi^{\bf 1}=\frac{1}{168}\chi^{(4)}_{ijkm} \star\varphi^{ijkm}~,~~~\chi^{\bf 7}_{i}=\frac{1}{6} \chi^{(4)}_{ijkm} \varphi^{jkm}~,~~~
\cr
&&
\chi^{\bf 27}_{ij}=-\frac{1}{3} \chi^{(4)}_{kmn(i} \star\varphi^{\phantom(}_{j)_o}{}^{kmn}~.
\ey
Using these definitions the solution of the linear system can be read off from appendix~\ref{linsyssol}.

There are many ways to arrange the solution of the linear system in appendix~\ref{g2sol}\@. The arrangement that we follow is to express the components
of the forms with higher degree, like $G$ and $H$,
in terms of the components of the forms of lower degree and the geometry.
In particular for the $G$ fluxes we have the following.
The $G_{-+1}^{\bf 7}$ component has been determined in (\ref{gbonex}). The $G_{-+}^{\bf 7}$ and $G_{-+}^{\bf 14}$ components have been given
in (\ref{gbtwox}) and (\ref{gplusminusf}),  respectively. $G_{-1}^{\bf 7}$ is described in (\ref{gminustwo}), while $G_{-1}^{\bf 14}$ is not restricted
by the KSEs.  Moreover, $G_{+1}^{\bf 7}$ vanishes according to (\ref{vplus}), while $G_{+1}^{\bf 14}$ is given in (\ref{restplus}).
Next $G_-^{\bf 1}$ and $G_-^{\bf 7}$ are presented in (\ref{gsminus}) and (\ref{gminusthree}), respectively, while $G_-^{\bf 27}$ is not restricted
by the KSEs. $G_+^{(3)}$ vanishes as can be seen from (\ref{restplus}). Furthermore $G^{\bf1}_1$, $G^{\bf7}_1$ and $G^{\bf 27}_1$ are given in (\ref{gones}),
(\ref{gatwox}) and (\ref{gst}), respectively. Similarly, the three components of $G^{(4)}$ are presented in (\ref{gfs}), (\ref{gaonexp}) and (\ref{gsf}),
respectively.

Next consider the $H$ fluxes. The $H_{-+1}$ component is given in (\ref{hpmo}). The component $H_{+-i}$ is described in (\ref{hpms}), $H_{+1i}$ vanishes according to (\ref{vplus}) and
$H_{-1i}$ is not restricted by the KSEs. $H_+^{\bf 7}$ vanishes according to (\ref{vplus}) while $H_+^{\bf 14}$ is given in (\ref{restplus}).
$H_-^{(2)}$ is not restricted by the KSEs. $H_1^{\bf 7}$ is presented in (\ref{honetwos}) while $H_1^{\bf 14}$ has been determined in (\ref{honef}).
Moreover the ${\bf 1}$ and ${\bf 7}$ components of $H^{(3)}$ are given in (\ref{hthreeo}) and (\ref{hthrees}), respectively,  while ${\bf 27}$ is not restricted by the KSEs.

It remains to describe the solution for the $F$ fluxes. $F_{-+}$ is given in (\ref{fmp}), $F_{+1}$ vanishes according to (\ref{splus}) and $F_{-1}$ is not restricted by the KSEs.
$F_{+i}$ vanishes, see (\ref{vplus}), $F_{1i}$ is determined by (\ref{dphicx}) and (\ref{dphidx}), and $F_{-i}$ is not restricted by the KSEs.  Moreover,
the ${\bf 7}$ component of $F^{(2)}$ is given by (\ref{dphicx}) and (\ref{dphidx}) while the ${\bf 14}$ component is not restricted by the KSEs.
Furthermore observe that the dilaton $\Phi$ as well as the functions $f,g$ are invariant under the vector field $K$ associated with $e^-$. It is expected that
all fields are invariant under the action of $K$.

Finally, the condition (\ref{fgconb}) can either be seen as a condition that restricts the dilaton or as a condition on the geometry of the spacetime. The latter case will be investigated below.

\subsection{Geometry of \texorpdfGtwoReight{} backgrounds}
\label{secgeo}

The conditions on the geometry of the spacetime that arise from the solution of the KSEs are given in the equations (\ref{plusgeom}), (\ref{geomconx}), the first condition
in (\ref{hpms}), (\ref{fgcon}) and (\ref{fgconb}).
After some investigation, the conditions (\ref{plusgeom}), (\ref{geomconx})
and (\ref{hpms}) can be re-expressed as in equation (\ref{geomg2}) of the introduction. Therefore the spacetime admits a null Killing vector field $K$ which
leaves the Killing spinor $\epsilon$ invariant.

Furthermore the rotation of $K$ is restricted.  For this observe that the sub-bundle $\mathrm{Ker} \, K\subset T^*M$ further splits as $\mathrm{Ker}\, K= I\oplus L$, where $I$ is spanned by $e^1$.  In addition,
 $\Lambda^2(L)$  decomposes in $G_2$ representations as  $\Lambda^2(L)=\Lambda'_{\bf 7}\oplus \Lambda_{\bf 7}\oplus\Lambda_{\bf 14}$ which have typical fibres
 $\mathbb{R}^7$,  $\mathbb{R}^7$ and $\mathfrak{g}_2$, respectively. The rotation $\rd e^-$ is a section of  $\Lambda'_{\bf 7}\oplus I \oplus \Lambda_{\bf 14}(L)$.  In particular
 \by
 \rd e^-= \rd e^-_{-i} e^-\wedge e^i+\rd e^-_{-1} e^-\wedge e^1+ \frac{1}{2} \rd e^-_{ij} e^i\wedge e^j~,~~~\rd e^-_{ij}\varphi^{ij}{}_k=0~.
 \ey
 Therefore the solution satisfies a $\mathfrak{g}_2$ instanton condition with an additional null component.

It remains to investigate the remaining two conditions (\ref{fgcon}) and (\ref{fgconb}).  Both these conditions are scalar conditions.
The former condition puts a restriction on the $G_2$ structure of the spacetime.  In fact, it restricts the singlet part of $\rd\varphi$.
In particular,  (\ref{fgcon})  can be rewritten as
\by
&&(f^{-1} g^{-1}-2 fg) \partial_1 (f^2-g^2)-\frac{1}{24}  fg \,\rd\varphi_{ijkm} \star\varphi^{ijkm}
+2 f g (f^2-g^2) \rd e^-_{-1}=0~.
\label{fgconxx}
\ey
The latter condition can also be interpreted as a restriction on the dilaton $\Phi$. As a geometric condition it can be rewritten as
\by
3 \partial_1\Phi -  \partial_1 \log (fg)  + \frac{1}{6} \nabla_i \varphi_{1jk} \varphi^{ijk} +2  \rd e^-_{-1} + 2 f g S=0~.
\label{fgconbxx}
\ey
Again it restricts the singlet class in the decomposition of $\nabla_i \varphi_{1jk}$ in terms of $G_2$ representations.

\section{Special case (\texorpdfstring{$f=g$}{f = g})}
\label{secspecial}

A special case arises whenever the 1-form bilinear $\kappa(\epsilon,\tilde\epsilon)$ vanishes, see appendix~\ref{sforms}\@. As this bilinear by construction is covariant, such a condition
can be implemented covariantly over the whole spacetime. In the gauge that we are working this implies that $f^2=g^2$ and so $f=\pm g$.  In such  case, the solution of the linear system simplifies dramatically. We shall take $f=g=1/\sqrt{2}$.
The other case can be treated symmetrically.

\subsection{Fluxes}

As the expression of the fluxes in terms of the geometry is relatively simple, we shall present the fluxes in a closed form.
In particular, we have that the 4-form flux is
\by
G&=&\big(-\frac{2}{5} \rd e^-_{i-}+\frac{1}{10} F_{mn} \varphi_i{}^{mn}+\frac{4}{5} \rd e^1_{1i}-\frac{1}{10} \varphi^{kmn} \nabla_k\varphi_{imn}\big) e^-\wedge e^+\wedge e^1\wedge e^i
\cr
&& +\big(\frac{1}{12} \rd e^1_{mn}\varphi^{mnp} \varphi_{pij}  +\frac{1}{2} (\rd e^1_{ij})^{\bf 14}\big) e^-\wedge e^+\wedge e^i\wedge e^j
\cr
&& +\frac{1}{2} (\rd e^-_{ij})^{\bf 14} e^+\wedge e^1\wedge e^i\wedge e^j
\cr
&& +\frac{1}{12} \big(-2 F_{-p}+2 H_{-1p}-\frac{1}{6} \nabla_-\varphi_{kmn} \star\varphi_p{}^{kmn}\big) \varphi^p{}_{ij} e^-\wedge e^1\wedge e^i\wedge e^j
+ e^-\wedge e^1\wedge G^{\bf 14}_{-1}
\cr
&&+\frac{1}{6} \big(-\frac{1}{10} F_{mn} \varphi_p{}^{mn}+\frac{2}{5} \rd e^-_{p-}+\frac{1}{5} \rd e^1_{1p}-\frac{1}{40} \varphi^{kmn} \nabla_k\varphi_{pmn}\big) \star\varphi^p{}_{ijk} e^1\wedge e^i\wedge e^j\wedge e^k
\cr
&&
+\frac{1}{48}\star\varphi_{(i}{}^{kmn} \nabla_{p)_o} \varphi_{kmn} \varphi^p{}_{jk} e^1\wedge e^i\wedge e^j\wedge e^k
\cr
&&+ \frac{1}{7} F_{-1} e^- \wedge\varphi-\big(\frac{1}{48} H_{-mn} \varphi_p{}^{mn}\star\varphi^p{}_{ijk}+\frac{1}{12} \nabla_-\star\varphi_{1ijk}\big)  e^-\wedge e^i\wedge e^j\wedge e^k+ e^-\wedge G_-^{\bf 27}
\cr
&&
+\big(-\frac{1}{7}S+\frac{2}{7} \rd e^-_{-1}\big) \star\varphi +\frac{1}{4! \, 12} \nabla_1\varphi_{pqt} \star\varphi_{i_1}{}^{pqt} \varphi_{i_2i_3i_4} e^{i_1}\wedge e^{i_2}\wedge e^{i_3}\wedge e^{i_4}
\cr
&&
-\frac{1}{4!} \big(H_{(p}{}^{mn} \varphi_{i_1)_o mn} +\frac{2}{3} \nabla_{(p }\varphi_{|1mn|} \varphi_{i_1)_o}{}^{mn}\big) \star\varphi^p{}_{i_2i_3i_4} e^{i_1}\wedge e^{i_2}\wedge e^{i_3}\wedge e^{i_4}~,
\ey
the 3-form flux is
\by
H&=& e^-\wedge H_-^{(2)}+ \frac{1}{30} \big(\frac{1}{2}\varphi^{kmn} \nabla_k\varphi_{pmn}+ 2 F_{mn} \varphi_p{}^{mn}-3 \rd e^-_{p-}
+6 \rd e^1_{1p}\big) \varphi^p{}_{ij} e^1\wedge e^i\wedge e^j
\cr
&&
-e^1\wedge F^{\bf 14}+ e^-\wedge e^1\wedge
H_{-1}^{(1)}
+\frac{2}{21} \big(2S-\frac{1}{6} \nabla_m \varphi_{1nq} \varphi^{mnq}-\frac{1}{2} \rd e^-_{-1}\big) \varphi
\cr
&&
+\frac{1}{6} \big(-\frac{1}{4} \rd e^1_{mn}\varphi_p{}^{mn}+ \frac{1}{48}  \nabla_1 \varphi_{mnq} \star\varphi_p{}^{mnq}\big) \star\varphi^p{}_{ijk} e^i\wedge e^j\wedge e^k
\cr &&  ~~~~+ H^{\bf 27}~,
\ey
and the 2-form flux is
\by
F&=&F_{-1} e^-\wedge e^1+ F_{-i} e^-\wedge e^i+\big(\nabla^p\varphi_{1pi}-\frac{1}{12} \nabla_1 \varphi_{mnq} \star\varphi_i{}^{mnq}\big) e^1\wedge e^i
\cr
&&
+\frac{1}{6}\big( 5 \rd\Phi_p-3 \rd e^-_{p-}+\frac{1}{2} \varphi^{kmn} \nabla_k\varphi_{pmn} + \rd e^1_{1p}\big) \varphi^p{}_{ij} e^i\wedge e^j+ F^{\bf 14}~.
\ey
Notice that the KSEs do not restrict all components of the fluxes in terms of the geometry.  In fact, there are several components
that remain unrestricted. Of course, these are  determined by the field equations and Bianchi identities of the theory.

\subsection{Geometry}

The restrictions on the geometry of the backgrounds  imposed by supersymmetry are as those of the general case described in (\ref{geomg2}).
Moreover as $f$ and $g$ are constant, there is  a simplification of the additional condition (\ref{fgconxx})  which can now be written
as in (\ref{geombg2}) which is the vanishing of the ${\bf 1}$ representation of  $\rd\varphi$. It is well-known  that this is one of the Gray-Hervella type of classes
for $G_2$ structures \cite{stefan}.  Therefore the $G_2$ structure of the spacetime is of restricted type.

The condition (\ref{fgconbxx}) also simplifies and in the special case reads as
\by
3 \partial_1\Phi   + \frac{1}{6} \nabla_i \varphi_{1jk} \varphi^{ijk} +2  \rd e^-_{-1} +  S=0~.
\label{fgconbxxx}
\ey
Again this condition can either be thought as a condition on the geometry or as an equation which determines the dilaton
in terms of the geometry.

\section{Concluding Remarks}
\label{secconclusion}

We have shown that the existence of a $G_2\ltimes \mathbb{R}^8$ Killing spinor on (massive) IIA backgrounds imposes rather weak conditions
on the geometry. In particular, the backgrounds are required to admit a null Killing vector which leaves the Killing
spinor invariant, and the rotation of the Killing vector is required to satisfy a $G_2$ instanton condition with an additional non-vanishing null component.
There are two additional scalar conditions  given in
(\ref{fgconxx}) and (\ref{fgconbxx}) which thus do not impose strong conditions on the geometry.
It is remarkable that these conditions on the geometry are similar to the ones we have found  for (massive) IIA backgrounds admitting $\SU(4)$ invariant Killing spinors \cite{iiaII}. In particular, $\SU(4)$ backgrounds   admit a time-like Killing vector field which also leaves the Killing spinor invariant.  The $\SU(4)$ backgrounds also admit another vector field
which commutes with the Killing vector field, and impose some additional scalar conditions on the geometry.
However the geometry of IIA backgrounds admitting either $\Spin(7)$ or  $\Spin(7)\ltimes \mathbb{R}^8$ invariant Killing spinors
satisfy stronger conditions \cite{iiaI}.  This is because  apart from the existence of appropriate Killing vector fields which leaves
 the Killing spinor invariant an additional condition, which transforms as a spinor under $\Spin(7)$, is required.

 Another characteristic of $\Spin(7)$, $\SU(4)$ and $G_2\ltimes \mathbb{R}^8$ backgrounds is the existence of  special cases for which the
solution of the KSEs simplify dramatically.  All three special cases arise as a requirement for the vanishing of a certain
spinor bilinear. As all these transform covariantly on the spacetime, these vanishing conditions can be imposed consistently over the whole spacetime.
The existence of such special spinors is reminiscent of the existence of pure spinors in Euclidean signature spaces. Pure spinors are complex and can also be  defined as those for which a certain vector bilinear vanishes.   As pure spinors are related to complex structures,
there may be a similar geometric interpretation for the Killing spinors that occur in all three special cases; though of course in IIA the spinors are real and it is not in all cases a vector bilinear that is required to vanish.

 As we have already
mentioned in the introduction, since  the rank of the spinor bundle of  IIA backgrounds is much larger than the dimension of the spacetime, the mere existence of
a no-where vanishing spinor does not imply the reduction of the structure group of the spin bundle.  As a result G-structures are not useful for the global
description of the geometry of spacetime.  This is also the case even when one adapts
  a generalized geometry approach to  characterize the backgrounds in terms of the isotropy groups
of the components of the Killing spinors in the  two different sectors.  In the context of IIA theory, this is to characterize the backgrounds in terms of the
isotropy groups of chiral and anti-chiral components of the Killing spinors. In the case of one supersymmetry, these are $\Spin(7)\ltimes \mathbb{R}^8 \times \Spin(9,1)$ for a $\Spin(7)\ltimes  \mathbb{R}^8$-invariant
Killing spinor, $\Spin(7)\ltimes \mathbb{R}^8\times \Spin(7)\ltimes \mathbb{R_*}^8$ for a $\Spin(7)$-invariant Killing spinor, $\Spin(7)\ltimes \mathbb{R}^8\times \SU(4)\ltimes \mathbb{R_*}^8$ for an $\SU(4)$-invariant Killing spinor, and $\Spin(7)\ltimes \mathbb{R}^8\times \Spin(7)\ltimes \mathbb{R}^8$ for a $G_2\ltimes  \mathbb{R}^8$-invariant Killing spinor, where $\mathbb{R_*}^8$ denotes the other lightcone direction compared to that of $\mathbb{R}^8$.
However, such a characterization may work well in the special cases where the two chiral components never vanish, like that of the common sector, but
for generic IIA backgrounds this may not always be the case.  So such a characterization is not useful in general type II theories.

Spinorial geometry \cite{spingeom}, on the other hand, can be adapted to treat all cases simultaneously. For this it suffices to choose a Killing spinor representative
which includes all the Killing spinors with the four different isotropy groups as special cases. There are  examples already of such a treatment.  The first example is the IIA $\Spin(7)$ backgrounds which
include those with $\Spin(7)\ltimes \mathbb{R}^8$ invariant Killing spinors as a special case. Another example is the Killing spinors of $G_2\ltimes \mathbb{R}^8$
 backgrounds which include the Killing spinors of the $\Spin(7)\ltimes \mathbb{R}^8$ backgrounds as special cases whenever $f$ or $g$ vanishes. In the general case, the construction of the linear system will be straightforward.  However, it will be more involved than  the ones we have considered so far  for two  reasons. First, it will depend
on many more functions like   $f$ and $g$ in (\ref{ksg2}), and second, as the manifest symmetry of the Killing spinor representative will be much smaller
compared to the backgrounds we have investigated here,  the linear system will decompose into many more equations. As a result, the solution will be more difficult to find, but since we have a computer implementation of the analysis this is not a problem.
In any case, spinorial geometry provides a method to solve the KSEs of  backgrounds for which the isotropy group of the Killing spinor changes from patch to patch,
as has already been demonstrated in the examples mentioned above.
As the change of isotropy group of the Killing spinor is expected to be a widespread phenomenon, for example all spacetimes which exhibit a Killing horizon
with respect to the Killing vector bilinear belong to this category, it will be fruitful to pursue this in the future.


\acknowledgments
UG is supported by the Knut and Alice Wallenberg Foundation and by the Swedish Research Council. GP is partially supported by the STFC grant  ST/L000326/1.

\appendix

\section{Conventions and useful formulae}
\label{conv}

Our IIA supergravity conventions including the form of the KSEs, Bianchi identities and field equations are those of \cite{iiaI}, which are similar to those of \cite{Bergshoeff:2010mv}.
Our spinor conventions can be found in \cite{iib1}.

\subsection{\texorpdfGtwo{} fundamental forms}

The linear system that arises from  the KSEs in the context of spinorial geometry is initially expressed in terms of $\SU(3)$ representations which is the
 manifest symmetry of the Killing spinor (\ref{ksg2}). However, the linear system can be  re-expressed in terms of $G_2$ representations, which is the isotropy group of the Killing spinor  (\ref{ksg2}).  This requires the use of the fundamental $G_2$  3-form
\bn
\varphi=\frac{1}{3!} \varphi_{i_1i_2i_3} e^{i_1}\wedge e^{i_2}\wedge e^{i_3}= {\rm Re \ } \chi +e^6 \wedge \omega~,
\en
and its 4-form dual
\bn
\star \varphi = - e^6\wedge {\rm Im \ } \chi - \frac{1}{2}\omega\wedge\omega~,
\en
where
\[
\omega = -(e^2 \wedge e^7 + e^3 \wedge e^8+ e^4 \wedge e^9) \notag~,~~\chi = (e^2 + i e^7)\wedge(e^3 + i e^8)\wedge(e^4 + i e^9)~,
\]
and the Hodge operator is taken with respect to the volume form $e^2\wedge e^3\wedge e^4\wedge e^6 \wedge e^7 \wedge e^8 \wedge e^9$.
As the initial formulation is in terms of $\SU(3)$ representations, we need expressions for the fundamental forms of $G_2$ in terms of
the fundamental forms of $\SU(3)$.

In the Hermitian basis,
\by
e^\alpha = \frac{1}{\sqrt{2}} (e^a + i e^{a+5}) ~,~~ e^{\bar \alpha} = \frac{1}{\sqrt{2}} (e^a - i e^{a+5}) ~,~~a=2,\ldots, 4~,
\ey
which arises in the initial construction of the linear system. We also have
\by
 \omega_{\alpha \bar \beta} = - i \delta_{\alpha \bar \beta} ~,~~\chi_{\alpha_1 \alpha_2 \alpha_3 } =2\sqrt{2} \epsilon_{\alpha_1 \alpha_2 \alpha_3 } ~,
\ey
and
\by
&\varphi_{\alpha\beta\gamma}=\sqrt{2}\epsilon_{\alpha\beta\gamma}~, ~~~~\varphi_{1 \beta\bar\gamma} = -\tfrac{1}{\sqrt{2}} \delta_{\beta \bar\gamma}~,\\
&(\star\varphi)_{\alpha\bar\beta\gamma\bar\delta}=\delta_{\alpha\bar\beta} \delta_{\gamma\bar\delta}-\delta_{\gamma\bar\beta} \delta_{\alpha\bar\delta}~,~~(\star\varphi)_{1 \alpha_1 \alpha_2 \alpha_3} = \epsilon_{\alpha_1 \alpha_2 \alpha_3} ~,~~(\star\varphi)_{\bar 1 \alpha_1 \alpha_2 \alpha_3} = -\epsilon_{\alpha_1 \alpha_2 \alpha_3}~,
\ey
where $1$ and $\bar 1$ are holomorphic and anti-holomorphic indices, respectively.
Using these formulae, one can easily find the various contractions of $\varphi$ and $\star \varphi$ as
\by
\varphi_{i l_1 l_2 }\,\varphi^{j l_1 l_2 }  &= & 6 \delta_{i}^{j} ~,\\
\varphi_{i_1 i_2 l }\,\varphi^{j_1 j_2 l}  &=& 2 \delta_{[i_1 i_2]} ^{\,j_1 j_2}- (\star\varphi)_{i_1 i_2}{}^{ j_1 j_2}  ~,\\
\varphi_{i}{}^{l_1 l_2}\, (\star\varphi)_{j_1 j_2 l_1 l_2} &=& -4 \varphi_{i j_1 j_2}~,\\
\varphi_{i_1 i_2 l}\, (\star\varphi)^{j_1 j_2 j_3 l} &=& 6 \delta_{[i_1}^{[j_1}\varphi_{i_2]}^{j_2 j_3]}~,\\
(\star\varphi)_{i l_1 l_2 l_3}\, (\star\varphi)^{j l_1 l_2 l_3} &=& 24 \delta_i^j~,\\
(\star\varphi)_{i_1 i_2 l_1 l_2 }\, (\star\varphi)^{j_1 j_2 l_1 l_2 } &=& 8 \delta_{[i_1 i_2]}^{j_1 j_2}-2 (\star\varphi)_{i_1 i_2}{}^{j_1 j_2}~,\\
(\star\varphi)_{i_1 i_2 i_3 l }\, (\star\varphi)^{j_1 j_2 j_3 l } &=& 6 \delta_{[i_1 i_2 i_3]}^{j_1 j_2 j_3} - 9 \delta_{[i_1}^{[j_1}(\star\varphi)_{i_2 i_3]}{}^{j_2 j_3]}- \varphi_{i_1 i_2 i_3}\varphi^{j_1 j_2 j_3}~,
\ey
which are useful in many of the computations we have performed,
where
\by
\delta_{j_1 j_2\dots j_n} ^{i_1 i_2\dots i_n} =\delta^{[i_1}{}_{[j_1} \delta^{i_2}{}_{j_2}\cdots\delta^{i_n]}{}_{[j_n]}~.
\ey
This summarizes $G_2$ fundamental form identities.

\subsection{Covariant derivatives of fundamental forms}

In the linear system, various components of the spin connection appear. Because of this it  seems that
the associated relations are not covariant.  However, this is not the case provided that the patching conditions
of the spacetime are compatible with the underlying $G_2$ structure. This is enforced by the observation
that all components of the spin connection that appear in the linear system  can be re-expressed as appropriate covariant derivatives
on the fundamental forms $\varphi$,  $\star\varphi$, $e^+, e^1$ and $e^-$.
In particular, we have
\by
\nabla_A \varphi_{B_1 B_2 B_3} &= &3 \Omega_{A,[B_1}{}^C \varphi_{|C| B_2 B_3 ]} ~,\\
\nabla_A (\star\varphi)_{B_1 B_2 B_3 B_4}& =& 4 \Omega_{A,[B_1}{}^C (\star\varphi)_{|C| B_2 B_3 B_4 ]} ~,
\ey
where $A,B,C$ are spacetime frame indices.
These give rise to the identities
\by
\star\varphi^{k_1 k_2 k_3 k_4}\nabla_{k_1} \varphi_{k_2  k_3 k_4} &=& -12 \Omega_{m,nq} \varphi^{mnq}~,\\
g^{A B_1} \varphi^{k_1 k_2 k_3} \nabla_A (\star\varphi)_{B_1 k_1 k_2 k_3}& =& 12 \Omega_{m,nq} \varphi^{mnq}~,\\
\varphi^{k_1 k_2 k_3}\nabla_{A} \varphi_{k_1  k_2 k_3} &=& 0~,\\
\varphi^{k_1 k_2 k_3}\nabla_{k_1} \varphi_{a  k_2 k_3} &=& 6 \Omega_{k, a}{}^k ~,\\
\varphi^{k_1 k_2 k_3}\nabla_{k_1} \varphi_{i  k_2 k_3} &=& 4 \Omega_{k, i}{}^k -2 \Omega_{m,nq} \star\varphi_{i}{}^{mnq}~,\\
\star\varphi^{k_1 k_2 k_3 k_4}\nabla_{A} \star\varphi_{k_1  k_2 k_3 k_4} &=& 0~,\\
\star\varphi^{k_1 k_2 k_3 k_4}\nabla_{k_1} \star\varphi_{a  k_2 k_3 k_4} &=& 24 \Omega_{k, a}{}^k ~,\\
\star\varphi^{k_1 k_2 k_3 k_4}\nabla_{k_1} \star\varphi_{i  k_2 k_3 k_4} &=& 12 \Omega_{k, i}{}^k -6 \Omega_{m,nq} \star\varphi_{i}{}^{mnq}~,\\
g^{A B_3} \nabla_A \varphi_{B_1 B_2 B_3} &= & -2\Omega_{m,n [B_1} \varphi_{B_2]}{}^{mn}+  \Omega_{C,}{}^{C k}\varphi_{B_1 B_2 k}~,\\
\star\varphi_{j}{}^{k_1 k_2 k_3}\nabla_{A} \varphi_{k_1 k_2  k_3 } &=& -12 \Omega_{A,m n} \varphi_j{}^{mn}~,\\
\star\varphi_{j}{}^{k_1 k_2 k_3}\nabla_{k_1} \varphi_{a k_2  k_3 } &=&  4 \Omega_{m, n a}\varphi_j{}^{mn}\\
\star\varphi_{j}{}^{k_1 k_2 k_3}\nabla_{k_1} \varphi_{i k_2  k_3 } &=& -2 \Omega_{i,m n}\varphi_j{}^{mn} + 2 \Omega_{m, n i}\varphi_j{}^{mn} + 2\Omega_{k,m}{}^k \varphi_{ij}{}^{m}\notag\\
&& - 2 \Omega_{m,n j} \varphi_{i}{}^{ m n} +2 g_{ij} \Omega_{m,nq} \varphi^{mnq}~,
\ey
where $a,b={-,+,1}$, i.e.~$A = \{a, i\}$, where $i$ is a seven-dimensional index labelling the $G_2$-directions.  We use the formulae above to express the conditions on the geometry in terms of the fundamental forms.


\section{The spinor bilinears of \texorpdfGtwoReight{} spinors}
\label{sforms}

The IIA spinors are Majorana, so for the computation of the bilinears one can use either the Majorana or Dirac inner products. The conventions for these can be found in  \cite{iib1}. In IIA supergravity, apart from the Killing spinor $\epsilon$ in (\ref{ksg2}), one can define another globally defined spinor  $\tilde \epsilon= \Gamma_{11}\epsilon$ which is not necessarily Killing. The form bilinears of  $\epsilon$ and $\tilde \epsilon$  are two 1-forms
\by
\kappa(\epsilon,\epsilon) = -(f^2+ g^2 ) e^-  ~,~~~
\kappa(\epsilon,\tilde\epsilon)=  (f^2- g^2 ) e^- ~,
\ey
a 2-form
\bn
\omega(\epsilon,\epsilon) = -2 f g e^- \wedge e^1~,
\en
a 4-form
\bn
\zeta(\epsilon,\tilde\epsilon) = 2 f g e^- \wedge \varphi~,
\en
and two 5-forms
\by
\tau(\epsilon,\epsilon) &=& -(f^2-g^2)e^-\wedge e^1\wedge \varphi -(f^2+g^2) e^-\wedge \star\varphi ~,\notag\\
\tau(\epsilon,\tilde\epsilon) &=&(f^2+g^2)e^-\wedge e^1\wedge \varphi +(f^2-g^2) e^-\wedge \star\varphi ~,
\ey
where $\varphi$ is the invariant $G_2$ 3-form defined in the previous appendix  and we have normalised the Killing spinor with an additional factor of $1/\sqrt{2}$.
It is convenient in many computations  to take the gauge $f^2+g^2=1$.

\section{The linear system for the \texorpdfGtwoReight{} invariant spinor}
\label{g2sol}

\subsection{Dilatino Killing spinor equation}

The linear system that arises from the KSEs in the context of spinorial geometry can be organized  in the scalar, vector (or equivalently ${\bf 7}$), 2-form, symmetric traceless (or equivalently ${\bf 27}$) $G_2$ representations. In particular, the conditions that arise from the dilatino KSE are as follows.

\subsubsection{Conditions in the scalar representation}

The conditions that transform as scalars are

\by
&f {\rm d}\Phi_+ +\tfrac{3}{4}g F_{+ 1} - \tfrac{1}{24} g G_{+ j_1 j_2 j_3} \varphi^{j_1 j_2 j_3}=0~,\label{algs1}\\
&g \,{\rm d}\Phi_+ -\tfrac{3}{4}f F_{+ 1} + \tfrac{1}{24} f G_{+ j_1 j_2 j_3} \varphi^{j_1 j_2 j_3}=0~,\label{algs2}\\
&f {\rm d}\Phi_1 +\tfrac{3}{4}g F_{-+}-\tfrac{1}{2} f H_{-+1}-\tfrac{1}{12} f H_{j_1 j_2 j_3} \varphi^{j_1 j_2 j_3}\qquad\qquad\notag\\
&\qquad\qquad-\tfrac{1}{24}g G_{1 j_1 j_2 j_3}\varphi^{j_1 j_2 j_3}+\tfrac{1}{96} g G_{j_1 j_2 j_3 j_4} (\star\varphi)^{j_1 j_2 j_3 j_4} +\tfrac{5}{4}g S=0~, \label{algs3}\\
&g {\rm d}\Phi_1 -\tfrac{3}{4}f F_{-+}+\tfrac{1}{2} g H_{-+1}-\tfrac{1}{12} g H_{j_1 j_2 j_3} \varphi^{j_1 j_2 j_3}\qquad\qquad\notag\\
&\qquad\qquad+\tfrac{1}{24}f G_{1 j_1 j_2 j_3}\varphi^{j_1 j_2 j_3}+\tfrac{1}{96} f G_{j_1 j_2 j_3 j_4} (\star\varphi)^{j_1 j_2 j_3 j_4} +\tfrac{5}{4}f S=0~. \label{algs4}
\ey
As we shall see later the solution of these conditions simplifies in the gauge $f^2+g^2=1$.

\subsubsection{Conditions in the vector representation}
The conditions that transform as vectors are
\by
-\tfrac{3}{2}g F_{+ i} - f H_{+ 1 i} - \tfrac{1}{12} g G_{+ j_1 j_2 j_3 } (\star\varphi)_i{}^{j_1 j_2 j_3} - \tfrac{1}{4} g G_{+ 1 j_1 j_2} \varphi_i{}^{j_1 j_2} +\tfrac{1}{2}f H_{+ j_1 j_2} \varphi_i{}^{j_1 j_2}=0,~\label{Aveceq1}\\
\tfrac{3}{2}f F_{+ i} + g H_{+ 1 i} - \tfrac{1}{12} f G_{+ j_1 j_2 j_3 } (\star\varphi)_i{}^{j_1 j_2 j_3} + \tfrac{1}{4} f G_{+ 1 j_1 j_2} \varphi_i{}^{j_1 j_2} +\tfrac{1}{2}g H_{+ j_1 j_2} \varphi_i{}^{j_1 j_2}=0,~\label{Aveceq2}\\
f {\rm d}\Phi_i-\tfrac{3}{4} g F_{1i} -\tfrac{3}{8}g F_{j_1 j_2}\varphi^{j_1 j_2}{}_i -\tfrac{1}{2 }f H_{-+i} +\tfrac{1}{4} f H_{1 j_1 j_2}\varphi^{j_1 j_2}{}_i  -\tfrac{1}{12} f H_{j_1 j_2 j_3} (\star\varphi)_i{}^{j_1 j_2 j_3}~~~\notag\\
 -\tfrac{1}{4} g G_{-+1i}-\tfrac{1}{8} g G_{- + j_1 j_2}\varphi_{i}{}^{j_1 j_2} -\tfrac{1}{24} g G_{i j_1 j_2 j_3} \varphi^{j_1 j_2 j_3} -\tfrac{1}{24}g G_{1 j_1 j_2 j_3} (\star\varphi)_i{}^{j_1 j_2 j_3}=0,~\label{Aveceq3}\\
g\,{\rm d}\Phi_i+\tfrac{3}{4} f F_{1i} -\tfrac{3}{8}f F_{j_1 j_2}\varphi^{j_1 j_2}{}_i +\tfrac{1}{2 }g H_{-+i} +\tfrac{1}{4} g H_{1 j_1 j_2}\varphi^{j_1 j_2}{}_i  +\tfrac{1}{12} g H_{j_1 j_2 j_3} (\star\varphi)_i{}^{j_1 j_2 j_3}~~~\notag\\
 -\tfrac{1}{4} f G_{-+1i}+\tfrac{1}{8} f G_{- + j_1 j_2}\varphi_{i}{}^{j_1 j_2} +\tfrac{1}{24} f G_{i j_1 j_2 j_3} \varphi^{j_1 j_2 j_3} -\tfrac{1}{24}f G_{1 j_1 j_2 j_3} (\star\varphi)_i{}^{j_1 j_2 j_3}=0.~\label{Aveceq4}
\ey
Note that the dilatino KSE gives rise to only scalar and vector conditions.

\subsection{Gravitino Killing spinor equation}

\subsubsection{Conditions in the symmetric traceless representation}

The conditions that lie in the ${\bf 27}$ representation are
\by
\tfrac{1}{4} g G_{+ (i }{}^{k_1 k_2}\varphi_{j)_o k_1 k_2} + f \Omega_{(i,j)_o+} = 0 ~, \label{dst1}\\
\tfrac{1}{4} f G_{+ (i }{}^{k_1 k_2}\varphi_{j)_o k_1 k_2} - g \Omega_{(i,j)_o+} = 0 ~,\label{dst2}\\
\tfrac{1}{12} f G_{(i }{}^{k_1 k_2 k_3} (\star\varphi)_{j)_o k_1 k_2 k_3}+\tfrac{1}{4}f G_{1 (i}{}^{k_1 k_2} \varphi_{j)_o k_1 k_2} +\tfrac{1}{4} g H_{(i}{}^{k_1 k_2}\varphi_{j)_o k_1 k_2}\qquad\qquad\notag\\
 - g \Omega_{(i,j)_o1} +\tfrac{1}{2} g \Omega_{(i,}{}^{k_1 k_2}\varphi_{j)_o k_1 k_2} =0~,\label{dst3}\\
- \tfrac{1}{12} g G_{(i }{}^{k_1 k_2 k_3} (\star\varphi)_{j)_o k_1 k_2 k_3}+\tfrac{1}{4}g G_{1 (i}{}^{k_1 k_2} \varphi_{j)_o k_1 k_2} -\tfrac{1}{4} f H_{(i}{}^{k_1 k_2}\varphi_{j)_o k_1 k_2}\qquad\qquad\notag\\
 + f \Omega_{(i,j)_o1} +\tfrac{1}{2} f \Omega_{(i,}{}^{k_1 k_2}\varphi_{j)_o k_1 k_2} =0~.\label{dst4}
\ey
As we shall show below these conditions impose restrictions on both  the geometry and fluxes of the theory.

\subsubsection{Conditions in the 2-form representation}
The conditions that arise from the gravitino KSE which are organized in the 2-form representation are
\by
\tfrac{1}{4} g G_{+ 1 i j} + \tfrac{1}{8} g G_{+ 1 }{}^{k_1 k_2} (\star\varphi)_{i j k_1 k_2} -\tfrac{1}{4}g F_{+}{}^k \varphi_{i j  k} +\tfrac{1}{2} f H_{+ i j} - f \Omega_{[i,j]+}= 0 ~,\label{Daseq1}\\
\tfrac{1}{4} f G_{+ 1 i j} + \tfrac{1}{8} f G_{+ 1 }{}^{k_1 k_2} (\star\varphi)_{i j k_1 k_2} -\tfrac{1}{4}f F_{+}{}^k \varphi_{i j  k} -\tfrac{1}{2} g H_{+ i j} - g \Omega_{[i,j]+}= 0 ~,\label{Daseq2}\\
f G_{-+i j} +\tfrac{1}{2} f G_{-+}{}^{ k_1 k_2} (\star\varphi)_{i j k_1 k_2} +f G_{- + 1}{}^k \varphi_{i j k} - f F_{i j }-\tfrac{1}{2} f F^{k_1 k_2}(\star\varphi)_{i j k_1 k_2}
\qquad\notag\\
-f F_{1}{}^k \varphi_{i j k} - 2 g H_{1 i j} + g H_{[i}{}^{k_1 k_2}\varphi_{j] k_1 k_2}-4 g\Omega_{[i,j]1} + 2 g \Omega_{[i,}{}^{k_1 k_2} \varphi_{j] k_1 k_2}  =0~, \label{Daseq3}\\
g G_{-+i j} +\tfrac{1}{2} g G_{-+}{}^{ k_1 k_2} (\star\varphi)_{i j k_1 k_2} -g G_{- + 1}{}^k \varphi_{i j k} + g F_{i j }+\tfrac{1}{2} g F^{k_1 k_2}(\star\varphi)_{i j k_1 k_2} \qquad\notag\\
-g F_{1}{}^k \varphi_{i j k}
 + 2 f H_{1 i j} + f H_{[i}{}^{k_1 k_2}\varphi_{j] k_1 k_2}-4 f\Omega_{[i,j]1} - 2 f \Omega_{[i,}{}^{k_1 k_2} \varphi_{j] k_1 k_2}  =0~. \label{Daseq4}
 \ey
The above conditions can be decomposed further into the ${\bf 7}$ and ${\bf 14}$ $G_2$ representations.  In particular the ${\bf 7}$ component
of (\ref{Daseq3}) and (\ref{Daseq4}) can be expressed as
\by
\label{7cond1}
&&-f G_{-+mn} \varphi^{mn}{}_i+ 6 f G_{-+1i}+f F_{mn} \varphi^{mn}{}_i- 6 f F_{1i}-2 g H_{1mn} \varphi^{mn}{}_i
\cr &&
+ g H_{mn\ell} \star\varphi^{mn\ell}{}_i- 4 g \Omega_{m,n1} \varphi^{mn}{}_i
+2 g \Omega_{m,n\ell} \star\varphi^{mn\ell}{}_i + 4 g \Omega_{k,i}{}^k =0~,
\ey
and
\by
\label{7cond2}
&&-g G_{-+mn} \varphi^{mn}{}_i- 6 g G_{-+1i}-g F_{mn} \varphi^{mn}{}_i- 6 g F_{1i}+2 f H_{1mn} \varphi^{mn}{}_i
\cr &&
+ f H_{mn\ell} \star\varphi^{mn\ell}{}_i- 4 f \Omega_{m,n1} \varphi^{mn}{}_i
-2 f \Omega_{m,n\ell} \star\varphi^{mn\ell}{}_i - 4 f \Omega_{k,i}{}^k=0~.
\ey
They will used later together with other  conditions in the vector representation to solve the linear system for the fluxes.

\subsubsection{Conditions in the vector representation}

The most involved set of conditions that arise in the linear system are those that lie  in the ${\bf 7}$ representation.  These are   as follows
\by
\Omega_{+,+ i} &=& 0 ,\label{Dveceq1}\\
H_{+ 1 i} + \Omega_{+,}{}^{k_1 k_2} \varphi_{i k_1 k_2} &=& 0 ,\label{Dveceq2}\\
\tfrac{1}{2} H_{+}{}^{k_1 k_2} \varphi_{i k_1 k_2} + 2 \Omega_{+,1 i} &=& 0, \label{Dveceq3}\\
\tfrac{1}{4}f G_{+ 1}{}^{k_1 k_2}\varphi_{i k_1 k_2} -\tfrac{1}{2} f F_{+ i} + g (\Omega_{i,+1}+\Omega_{1,+ i})  &=& 0 , \label{Dveceq4}\\
\tfrac{1}{4}g G_{+ 1}{}^{k_1 k_2}\varphi_{i k_1 k_2} -\tfrac{1}{2} g F_{+ i} - f (\Omega_{i,+1}+\Omega_{1,+ i}) &=& 0 , \label{Dveceq5}\\
\tfrac{1}{12} f G_{+}{}^{ k_1 k_2 k_3} (\star\varphi)_{i k_1 k_2 k_3} - g H_{+1i} - g( \Omega_{i,+1}-\Omega_{1,+ i}) &=& 0, \label{Dveceq6}\\
\tfrac{1}{12} g G_{+}{}^{ k_1 k_2 k_3} (\star\varphi)_{i k_1 k_2 k_3} + f H_{+1i} - f( \Omega_{i,+1}-\Omega_{1,+ i}) &=& 0,\label{Dveceq7}\\
{\text{d}f_i} -\tfrac{1}{48}g G_{1}{}^{ k_1 k_2 k_3}  (\star\varphi)_{i k_1 k_2 k_3} +\tfrac{1}{48} g G_{i}{}^{k_1 k_2 k_3} \varphi_{k_1 k_2 k_3}+ \tfrac{1}{8} g G_{-+1i} &&\notag \\
-\tfrac{1}{16} g G_{-+}{}^{k_1 k_2} \varphi_{i k_1 k_2} + \tfrac{1}{8} g F_{1i} -\tfrac{1}{16} g F^{k_1 k_2} \varphi_{i k_1 k_2}-\tfrac{1}{4} f H_{- + i}+ \tfrac{1}{2} f \Omega_{i,-+} &=&0 , \label{Dveceq8}\\
{\text{d}g_i} -\tfrac{1}{48}f G_{1}{}^{ k_1 k_2 k_3}  (\star\varphi)_{i k_1 k_2 k_3} -\tfrac{1}{48} f G_{i}{}^{k_1 k_2 k_3} \varphi_{k_1 k_2 k_3}+ \tfrac{1}{8} f G_{-+1i} &&\notag \\
+\tfrac{1}{16} f G_{-+}{}^{k_1 k_2} \varphi_{i k_1 k_2} - \tfrac{1}{8} f F_{1i} -\tfrac{1}{16} f F^{k_1 k_2} \varphi_{i k_1 k_2}+\tfrac{1}{4} g H_{- + i}+ \tfrac{1}{2} g \Omega_{i,-+} &=&0 ,\label{Dveceq9}\\
-\tfrac{1}{48}f G_{1}{}^{ k_1 k_2 k_3}  (\star\varphi)_{i k_1 k_2 k_3} -\tfrac{1}{48} f G_{i}{}^{k_1 k_2 k_3} \varphi_{k_1 k_2 k_3}+ \tfrac{1}{8} f F_{1i}-\tfrac{1}{16} f G_{-+}{}^{k_1 k_2} \varphi_{i k_1 k_2}  &&\notag \\
- \tfrac{1}{8} f G_{-+1i} +\tfrac{1}{16} f F^{k_1 k_2} \varphi_{i k_1 k_2}-\tfrac{1}{8}g H_{1}{}^{k_1 k_2} \varphi_{i k_1 k_2}- \tfrac{1}{2} g \Omega_{1,1i} -\tfrac{1}{4} g \Omega_{1,}{}^{k_1 k_2}\varphi_{i k_1 k_2} &=&0 , \label{Dveceq10}~~~\\
-\tfrac{1}{48}g G_{1}{}^{ k_1 k_2 k_3}  (\star\varphi)_{i k_1 k_2 k_3} +\tfrac{1}{48} g G_{i}{}^{k_1 k_2 k_3} \varphi_{k_1 k_2 k_3}- \tfrac{1}{8} g F_{1i}+\tfrac{1}{16} g G_{-+}{}^{k_1 k_2} \varphi_{i k_1 k_2}  &&\notag \\
- \tfrac{1}{8} g G_{-+1i} +\tfrac{1}{16} g F^{k_1 k_2} \varphi_{i k_1 k_2}-\tfrac{1}{8}f H_{1}{}^{k_1 k_2} \varphi_{i k_1 k_2}- \tfrac{1}{2} f \Omega_{1,1i} +\tfrac{1}{4} f \Omega_{1,}{}^{k_1 k_2}\varphi_{i k_1 k_2} &=&0 ,~~~\label{Dveceq11}\\
-\tfrac{1}{48}f G_{1}{}^{ k_1 k_2 k_3}  (\star\varphi)_{i k_1 k_2 k_3} +\tfrac{1}{48} f G_{i}{}^{k_1 k_2 k_3} \varphi_{k_1 k_2 k_3}-\tfrac{1}{16} f G_{-+}{}^{k_1 k_2} \varphi_{i k_1 k_2}  &&\notag \\
+ \tfrac{1}{8} f G_{-+1i} + \tfrac{1}{8} f F_{1i}-\tfrac{1}{16} f F^{k_1 k_2} \varphi_{i k_1 k_2}+\tfrac{1}{4} g H_{-+i}+ \tfrac{1}{2} g \Omega_{-,+i}  &=&0 ,~~~\label{Dveceq12}\\
\tfrac{1}{48}g G_{1}{}^{ k_1 k_2 k_3}  (\star\varphi)_{i k_1 k_2 k_3} +\tfrac{1}{48} g G_{i}{}^{k_1 k_2 k_3} \varphi_{k_1 k_2 k_3}-\tfrac{1}{16} g G_{-+}{}^{k_1 k_2} \varphi_{i k_1 k_2}  &&\notag \\
- \tfrac{1}{8} g G_{-+1i} + \tfrac{1}{8} g F_{1i}+\tfrac{1}{16} g F^{k_1 k_2} \varphi_{i k_1 k_2}+\tfrac{1}{4} f H_{-+i}- \tfrac{1}{2} f \Omega_{-,+i}  &=&0 ,~~~\label{Dveceq13}\\
-\tfrac{1}{6}f G_{-}{}^{ k_1 k_2 k_3}  (\star\varphi)_{i k_1 k_2 k_3} +\tfrac{1}{2} f G_{-1 }{}^{k_1 k_2 } \varphi_{i k_1 k_2 }+ f F_{-i}- g H_{-1i} &&\notag\\
-\tfrac{1}{2} g H_{-}{}^{ k_1 k_2} \varphi_{i k_1 k_2}-2  g \Omega_{-,1i} - g \Omega_{-,}{}^{k_1 k_2} \varphi_{i k_1 k_2}  &=&0 ,~~~\label{Dveceq14}\\
+\tfrac{1}{6}g G_{-}{}^{ k_1 k_2 k_3}  (\star\varphi)_{i k_1 k_2 k_3} +\tfrac{1}{2} g G_{-1 }{}^{k_1 k_2 } \varphi_{i k_1 k_2 }+ g F_{-i}- f H_{-1i} &&\notag\\
+\tfrac{1}{2} f H_{-}{}^{ k_1 k_2} \varphi_{i k_1 k_2}+2  f \Omega_{-,1i} - f \Omega_{-,}{}^{k_1 k_2} \varphi_{i k_1 k_2}  &=&0 .~~~\label{Dveceq15}
\ey
The solution of these conditions, together with the vector representation conditions that we derived  from the 2-form representation above and those that we have
found  from the dilatino KSE, gives a large number of relations between the components of the fluxes as well as conditions on the geometry of spacetime.

\subsubsection{Conditions in the scalar representation}
Finally, the conditions that arise in the scalar representation of $G_2$ are
\by
\Omega_{+,+ 1} &=& 0 ,\label{Dseq1}\\
{\text{d}f_+} +\tfrac{1}{2} f \Omega_{+,-+} &=& 0 ~,\label{Dseq2}\\
{\text{d}g_+} +\tfrac{1}{2} g \Omega_{+,-+} &=& 0 ~,\label{Dseq3}\\
{\text{d}f_-} +\tfrac{1}{24} g G_{-}{}^{k_1 k_2 k_3} \varphi_{k_1 k_2 k_3} -\tfrac{1}{4} g F_{- 1}+\tfrac{1}{2} f \Omega_{-,-+} &=& 0 ~,\label{Dseq4}\\
{\text{d}g_-} -\tfrac{1}{24} f G_{-}{}^{k_1 k_2 k_3} \varphi_{k_1 k_2 k_3} +\tfrac{1}{4} f F_{- 1}+\tfrac{1}{2} g \Omega_{-,-+} &=& 0 ~,\label{Dseq5}\\
f F_{+1} +\tfrac{1}{2} g \Omega_{1,1+} + \tfrac{1}{2} g \Omega_{k,+}{}^k &=& 0 ~, \label{Dseq6}\\
g F_{+1} -\tfrac{1}{2} f \Omega_{1,1+} - \tfrac{1}{2} f \Omega_{k,+}{}^k &=& 0 ~,\label{Dseq7}\\
f G_{+}{}^{k_1 k_2 k_3}\varphi_{k_1 k_2 k_3} + 21 g \Omega_{1,1+} - 3 g \Omega_{k,+}{}^k &=& 0 ~, \label{Dseq8}\\
g G_{+}{}^{k_1 k_2 k_3}\varphi_{k_1 k_2 k_3} - 21 f \Omega_{1,1+} + 3 f \Omega_{k,+}{}^k &=& 0 ~, \label{Dseq9}\\
{\text{d}f_1} + g F_{-+} -\tfrac{1}{8} f H^{k_1 k_2 k_3}\varphi_{k_1 k_2 k_3} - \tfrac{1}{4} f H_{- + 1}  +\tfrac{1}{4} f \Omega^{k_1}{}_{,}{}^{k_2 k_3} \varphi_{k_1 k_2 k_3}&&\notag\\
 -\tfrac{1}{2} f \Omega_{k,1}{}^k +\tfrac{1}{2} f \Omega_{1,-+}+ g S &=& 0 ~,\label{Dseq10}\\
 {\text{d}g_1} - f F_{-+} -\tfrac{1}{8} g H^{k_1 k_2 k_3}\varphi_{k_1 k_2 k_3} + \tfrac{1}{4} g H_{- + 1}  -\tfrac{1}{4} g \Omega^{k_1}{}_{,}{}^{k_2 k_3} \varphi_{k_1 k_2 k_3}&&\notag\\
 -\tfrac{1}{2} g \Omega_{k,1}{}^k +\tfrac{1}{2} g \Omega_{1,-+}+ f S &=& 0 ~,\label{Dseq11}\\
 \tfrac{1}{24} f G^{k_1 k_2 k_3 k_4} (\star\varphi)_{k_1 k_2 k_3 k_4} + 4 f F_{-+} +\tfrac{1}{2} g H^{k_1 k_2 k_3}\varphi_{k_1 k_2 k_3 } +g H_{-+1} &&\notag\\
+ g \Omega^{k_1}{}_{,}{}^{k_2 k_3} \varphi_{k_1 k_2 k_3} +2 g \Omega_{k,1}{}^k +2 g \Omega_{-,+1} - 3 f S &=& 0 ~, \label{Dseq12}\\
 \tfrac{1}{24} g G^{k_1 k_2 k_3 k_4} (\star\varphi)_{k_1 k_2 k_3 k_4} - 4 g F_{-+} +\tfrac{1}{2} f H^{k_1 k_2 k_3}\varphi_{k_1 k_2 k_3 } -f H_{-+1} &&\notag\\
- f \Omega^{k_1}{}_{,}{}^{k_2 k_3} \varphi_{k_1 k_2 k_3} +2 f \Omega_{k,1}{}^k +2 f \Omega_{-,+1} - 3 g S &=& 0 ~,\label{Dseq13}\\
\tfrac{1}{6} f G_{1}{}^{k_1 k_2 k_3} \varphi_{k_1 k_2 k_3} - 3 f F_{-+} -\tfrac{1}{2} g H^{k_1 k_2 k_3}\varphi_{k_1 k_2 k_3 } +g H_{-+1} &&\notag\\
- g \Omega^{k_1}{}_{,}{}^{k_2 k_3} \varphi_{k_1 k_2 k_3} -2 g \Omega_{k,1}{}^k +2 g \Omega_{-,+1} + 4 f S &=& 0 ~, \label{Dseq14}\\
\tfrac{1}{6} g G_{1}{}^{k_1 k_2 k_3} \varphi_{k_1 k_2 k_3} - 3 g F_{-+} +\tfrac{1}{2} f H^{k_1 k_2 k_3}\varphi_{k_1 k_2 k_3 } +f H_{-+1} &&\notag\\
- f \Omega^{k_1}{}_{,}{}^{k_2 k_3} \varphi_{k_1 k_2 k_3} +2 f \Omega_{k,1}{}^k -2 f \Omega_{-,+1} - 4 g S &=& 0 ~. \label{Dseq15}
\ey
The solution of these conditions, as well as all the others we have derived  from the dilatino and gravitino KSEs, will be given in the next appendix.

\section{Solution of the linear system}
\label{linsyssol}

\subsection{The + components of the fluxes and geometry }

Let us first consider the conditions on the $+$ components of the fluxes. After some computation and using the gauge $f^2+g^2=1$, the full content of the
conditions (\ref{algs1}), (\ref{algs2}), (\ref{Aveceq1}), (\ref{Aveceq2}), (\ref{Dveceq1}), (\ref{Dveceq2}), (\ref{Dveceq3}), (\ref{Dveceq4}),
(\ref{Dveceq5}), (\ref{Dveceq6}), (\ref{Dveceq7}), (\ref{dst1}), (\ref{dst2}), (\ref{Daseq1}), (\ref{Daseq2}), (\ref{Dseq1}), (\ref{Dseq2}), (\ref{Dseq3}),
(\ref{Dseq6}), (\ref{Dseq7}), (\ref{Dseq8}) and (\ref{Dseq9}) are the restrictions
\by
&&\Omega_{+,+1}=\Omega_{+,-+}=\Omega_{1,1+}=0 ~,\cr
&&\Omega_{+, 1 i} = \Omega_{1,+ i} = \Omega_{i,+1} =\Omega_{+,+i}=0~,~~~
\cr
&&\Omega_{+,}{}^{k_1 k_2} \varphi_{i k_1 k_2}  = 0 ~,~~~
  \Omega_{k_1,k_2 +} \varphi^{k_1 k_2}{}_i = 0 ~, ~~~
  \Omega_{(i,j)+}=0~,
  \label{plusgeom}
\ey
on the geometry, and the conditions
\by
F_{+1}=0~,~~\rd\Phi_+=0~,~~~\rd f_+=\rd g_+=0~,
\label{splus}
\ey
\by
F_{+i}=0~,~~~H_{+ij}\varphi^{ij}{}_k=0~,~~~H_{+1i}=0~,~~~G_{+1ij}\varphi^{ij}{}_k=0~,
\label{vplus}
\ey
\by
H_{+ij}=2 (f^2-g^2) \Omega_{i,j+}~,~~~G_{+1ij}=4fg \Omega_{i,j+}~,~~~G_{+ijk}=0~,
\label{restplus}
\ey
on the fluxes.  In particular, observe that most of these fluxes vanish.

\subsection{Solution of the rest of the scalar conditions}

The next task is to solve all the remaining scalar conditions of the linear system that
arises from the gravitino and dilatino KSEs in the context of spinorial geometry.  In particular
combining (\ref{Dseq4}) and (\ref{Dseq5}), we find
\by
&&\Omega_{-,- +} =0~,\label{geomconx}\\
&&\partial_- (f^2 - g^2) - f g F_{-1} + \frac{1}{6} f g G_{- i j k} \varphi^{i j k} = 0~. \label{gsminus}
\ey
Set $A=g (\ref{Dseq12})+ f (\ref{Dseq13})$, $B=g (\ref{Dseq12})- f (\ref{Dseq13})$, $C=g (\ref{Dseq14})+ f (\ref{Dseq15})$ and $D= g (\ref{Dseq14})- f (\ref{Dseq15})$, where
\by
&&D\equiv -\frac{1}{2} H_{ijk} \varphi^{ijk} + (g^2-f^2) H_{-+1}+ (f^2-g^2) \Omega_{i,jk} \varphi^{ijk}
\cr &&
-2 \Omega_{k,1}{}^k+ 2 \Omega_{-,+1}+ 8fg S=0~.
\ey
Then set
$C'=C+ (f^2-g^2) D$, where
\by
C'\equiv \frac{1}{3} G_{1ijk} \varphi^{ijk}+ 4 fg H_{-+1}- 4 fg \Omega_{i,jk} \varphi^{ijk}- 6 F_{-+}+ 8 (f^2-g^2) S=0~,
\ey
and consider $A'=A+D$ to find
\by
A'\equiv \frac{1}{12} fg G_{ijkl} \star \varphi^{ijkl}+ 2 (g^2-f^2) H_{-+1}+ 4 \Omega_{-,+1}+ 2 fg S=0~.
\ey
Next write $B'=B+(g^2-f^2) D$ to get
\by
&&B'\equiv 8fg F_{-+}+ (2-4 f^2 g^2) H_{-+1} + 4 f^2 g^2 \Omega_{i,jk} \varphi^{ijk}
\cr &&
- 4(f^2-g^2) \Omega_{-,+1}+ 8 fg (g^2-f^2) S=0~.
\ey
To continue, take $ f (\ref{Dseq10})+ g (\ref{Dseq11})$ and after using $D$  and $g^2+ f^2=1$, one finds
\by
\Omega_{-,+1}+\Omega_{1,+-}=0~.
\label{geomxxx}
\ey
Moreover, after eliminating the $H\varphi$ term,  $ f (\ref{Dseq10})- g (\ref{Dseq11})$ gives
\by
&&E\equiv f^{-1} g^{-1} \partial_1(f^2-g^2)+ 4 F_{-+}- 2 fg H_{-+1}+ 2 fg \Omega_{i,jk} \varphi^{ijk}
\cr
&&
-4 (f^2-g^2) S=0~.
\label{eee}
\ey
Furthermore, $f (\ref{algs3})+g (\ref{algs4})$ and $f (\ref{algs3})-g (\ref{algs4})$ yield
\by
&&\partial_1\Phi+\frac{1}{6} (f^2-g^2) H_{-+1}- \frac{1}{6} (f^2-g^2) \Omega_{i,jk} \varphi^{ijk}+ \frac{1}{3} \Omega_{k,1}{}^k
\cr &&
-\frac{4}{3} \Omega_{-,+1} +\frac{2}{3} fg S=0~,
\ey
and
\by
&&(f^2-g^2) \rd\Phi_1-\frac{1}{3} (1- f^2 g^2) H_{-+1}-\frac{1}{6} (1+2 f^2 g^2) \Omega_{i,jk} \varphi^{ijk}
\cr &&
+ \frac{1}{3} (f^2-g^2)( \Omega_{k,1}{}^k- \Omega_{-,+1})+\frac{2}{3} fg (f^2-g^2) S=0~,
\ey
respectively.

Observe that the system for $H_{-+1}$ and $F_{-+}$ components of the fluxes is over-determined so it may give some additional geometric constraints.
In particular, taking $B'-2fg E$, we get
\by
 H_{-+1}- \partial_1 (f^2-g^2)- 2 (f^2-g^2) \Omega_{-,+1}=0~,
 \label{hpmo}
\ey
which determines $H_{-+1}$. Putting this  back into (\ref{eee}), we find
\by
&&4 F_{-+}+ (f^{-1} g^{-1}-2 fg) \partial_1 (f^2-g^2)+2  fg  \Omega_{i,jk} \varphi^{ijk}
\cr
&&
- 4 f g (f^2-g^2) \Omega_{-,+1}- 4(f^2-g^2) S=0~,
\label{fnnpp}
\ey
which in turn specifies $F_{-+}$ in terms of geometry.
Substituting these expressions back into the previous equations, one gets
\by
&&-\frac{1}{2} H_{ijk} \varphi^{ijk}+ 2\partial_1 (f^2 g^2)+ 8 f^2 g^2 \Omega_{-,+1} + (f^2-g^2) \Omega_{i,jk} \varphi^{ijk}
\cr
&&~~~~~~~~~~- 2 \Omega_{k,1}{}^k+ 8 fg S=0~,
\label{hthreeo}
\ey
\by
\frac{1}{12} G_{ijkl} \star \varphi^{ijkl}+ 8 \partial_1 (f g)+ 16 f g \Omega_{-,+1} + 2 S=0~,
\label{gfs}
\ey
\by
\label{G13sol}
&&\frac{1}{3} G_{1ijk} \varphi^{ijk} + (\frac{3}{2} f^{-1} g^{-1} + fg) \partial_1 (f^2-g^2) + 2 fg (f^2-g^2) \Omega_{-,+1}
\cr
&&-fg \Omega_{ijk} \varphi^{ijk} + 2 (f^2-g^2) S=0~,
\ey
\by
\label{dilsol}
\partial_1\Phi-\frac{1}{3} \partial_1(f^2 g^2)-(1+\frac{4}{3} f^2 g^2) \Omega_{-,+1} -\frac{1}{6} (f^2-g^2) \Omega_{ijk} \varphi^{ijk}+\frac{1}{3} \Omega_{k,1}{}^k+\frac{2}{3} fg S=0,~~~
\ey
and
\by
&&(f^2-g^2) \partial_1 \Phi-\frac{1}{3} (1-f^2 g^2) \partial_1 (f^2-g^2)-\frac{1}{6} (1+2 f^2 g^2) \Omega_{ijk} \varphi^{ijk}
\cr
&&~~~~~~~~~
+(f^2-g^2) \Big(\frac{1}{3} \Omega_{k,1}{}^k+(\frac{2}{3} f^2 g^2-1)\Omega_{-,+1}\Big)
+\frac{2}{3}
fg (f^2-g^2) S=0~.
\ey
In particular after eliminating $\partial_1 \Phi$, the last equation can be rewritten as
\by
&&(f^{-1} g^{-1}-2 fg) \partial_1 (f^2-g^2)+2  fg  \Omega_{i,jk} \varphi^{ijk}
- 4 f g (f^2-g^2) \Omega_{-,+1}=0~,
\label{fgcon}
\ey
which is a constraint on the geometry.
Note also that using this geometric constraint, (\ref{fnnpp}) can be written as
\by
F_{-+}= (f^2-g^2) S~,
\label{fmp}
\ey
which determines a component of $F$ in terms of $S$.
Similarly, using (\ref{fgcon}) we can simplify two of the above expressions ((\ref{G13sol}) and (\ref{dilsol})), thus obtaining
\by
\frac{1}{6} G_{1ijk} \varphi^{ijk} + f^{-1} g^{-1} \partial_1 (f^2-g^2)+  (f^2-g^2) S = 0~,
\label{gones}
\ey
\by
3 \partial_1\Phi -  \partial_1 \log (fg)  + \Omega_{k,1}{}^k -4  \Omega_{-,+1} + 2 f g S=0~.
\label{fgconb}
\ey
The latter can either be interpreted as a geometric condition or used to express the dilaton $\Phi$ in terms of the geometry.

\subsection{Solution of the rest of vector conditions}

Considering $f(\ref{Dveceq8})\pm g (\ref{Dveceq9})$ and $ g (\ref{Dveceq12})\pm (\ref{Dveceq13})$ and recombining the results, we find
\by
&&\Omega_{i,+-}+\Omega_{-,+i}=0~,~~~\partial_i (f^2-g^2)-H_{-+i}+ 2 (f^2-g^2) \Omega_{i,-+}=0~,
\label{hpms}
\ey
\by
\frac{1}{24} fg G_{imnl} \varphi^{mnl}-\frac{1}{8} fg G_{+-mn} \varphi_i{}^{mn}+\frac{1}{4} fg F_{1i}+\frac{1}{4} \partial_i (f^2-g^2)=0~,
\label{gaonex}
\ey
\by
-\frac{1}{24}G_{1mnl} \star\varphi_i{}^{mnl}+\frac{1}{4} G_{-+1i}-\frac{1}{8} F_{mn} \varphi_i{}^{mn}+2 fg \Omega_{i,-+}+\partial_i(fg)=0~.
\label{gatwox}
\ey

Next taking $ g (\ref{Dveceq10})\pm f (\ref{Dveceq11})$ and using the above equations, we find
\by
&&-\frac{1}{2} fg G_{-+1i}- fg\partial_i (fg)-2 f^2g^2 \Omega_{i,-+}+\frac{1}{4} fg F_{mn}\varphi_i{}^{mn}-\frac{1}{8} H_{1mn} \varphi_i{}^{mn}-\frac{1}{2} \Omega_{1,1i}
\cr &&
+
\frac{1}{4} (f^2-g^2) \Omega_{1,mn} \varphi_i{}^{mn}=0~,
\label{gbonex}
\ey
\by
&&-\frac{1}{4} fg G_{-+mn} \varphi_i{}^{mn}+\frac{1}{2} fg F_{1i}+\frac{1}{4} \partial_i(f^2-g^2)+\frac{1}{8} (f^2-g^2) H_{1mn} \varphi_i{}^{mn}
\cr &&
+\frac{1}{2}
(f^2-g^2) \Omega_{1,1i}-\frac{1}{4} \Omega_{1,mn} \varphi_i{}^{mn}=0~.
\label{gbtwox}
\ey

Furthermore $g (\ref{7cond1})\pm f (\ref{7cond2})$ yield
\by
&&-2fg G_{-+mn} \varphi_i{}^{mn}-12 fg F_{1i}+ 2(f^2-g^2) H_{1mn} \varphi_i{}^{mn}-H_{mnl} \star\varphi_i{}^{mnl}-4 \Omega_{m,n1} \varphi_i{}^{mn}
\cr &&~~~~
+ 2(f^2-g^2) \Omega_{m,nl}
\star\varphi_i{}^{mnl}  - 4 (f^2 - g^2) \Omega_{k,i}{}^k=0~,
\label{gconex}
\ey
and
\by
&&12 fg G_{-+1i}+2 fg F_{mn} \varphi_i{}^{mn}-2 H_{1mn} \varphi_i{}^{mn}+ (f^2-g^2) H_{mnl}   \star\varphi_i{}^{mnl}
\cr &&~~~~
+ 4(f^2-g^2) \Omega_{m,n1} \varphi_i{}^{mn}
-2 \Omega_{m,nl}\star\varphi_i{}^{mnl} + 4  \Omega_{k,i}{}^k=0~.
\label{gctwox}
\ey

Also setting $g (\ref{Dveceq14})\pm f (\ref{Dveceq15})$, we have
\by
&&fg G_{-1mn} \varphi_i{}^{mn}+ 2fg F_{-i}- H_{-1i}+\frac{1}{2} (f^2-g^2) H_{-mn} \varphi_i{}^{mn}+ 2 (f^2-g^2) \Omega_{-,1i}
\cr
&&~~~~- \Omega_{-,mn} \varphi_i{}^{mn}=0~,
\label{gminustwo}
\ey
and
\by
&&-\frac{1}{3} fg G_{-mnl} \star\varphi_i{}^{mnl}+ (f^2-g^2) H_{-1i}-\frac{1}{2} H_{-mn} \varphi_i{}^{mn}-2 \Omega_{-,1i}
\cr &&
~~~~+ (f^2-g^2) \Omega_{-,mn} \varphi_i{}^{mn}=0~.
\label{gminusthree}
\ey

Then considering $f (\ref{Aveceq3})\pm g  (\ref{Aveceq4})$, we get
\by
&&\rd\Phi_i-\frac{3}{4}fg F_{mn} \varphi_i{}^{mn}-\frac{1}{2} (f^2-g^2) H_{-+i}+\frac{1}{4} H_{1mn} \varphi_i{}^{mn}
\cr &&
-\frac{1}{12} (f^2-g^2) H_{mnl}  \star\varphi_i{}^{mnl} -\frac{1}{2}fg G_{-+1i} -\frac{1}{12} fg
G_{1mnl}\star\varphi_i{}^{mnl}=0~,
\cr
&&
(f^2-g^2) \rd\Phi_i-\frac{3}{2} fg F_{1i}-\frac{1}{2} H_{-+i}+\frac{1}{4} (f^2-g^2) H_{1mn}\varphi_i{}^{mn}- \frac{1}{12} H_{mnl}   \star\varphi_i{}^{mnl}
 \cr
 &&
 -\frac{1}{4} fg G_{-+mn} \varphi_i{}^{mn}
-\frac{1}{12} fg G_{imnl} \varphi^{mnl}=0~.
\ey

Next take $(\ref{gaonex})-\frac{1}{2} (\ref{gbtwox})$ to get
\by
&&\frac{1}{24} fg G_{ikmn} \varphi^{kmn}+\frac{1}{8} \partial_i (f^2-g^2)-\frac{1}{16} (f^2-g^2) H_{1mn} \varphi_i{}^{mn}
\cr
&&
~~~~~~~~-\frac{1}{4} (f^2-g^2) \Omega_{1,1i}+\frac{1}{8}
\Omega_{1,mn} \varphi_i{}^{mn}=0~.
\label{gaonexp}
\ey
In addition  $fg (\ref{gatwox}+\frac{1}{2} (\ref{gbonex})$ gives
\by
&&-\frac{1}{24} fg G_{1kmn} \star\varphi_i{}^{kmn}+ f^2 g^2 \Omega_{i,-+} +\frac{1}{2} fg \partial_i(fg)-\frac{1}{16} H_{1mn} \varphi_i{}^{mn}
\cr
&&
~~~~~~~~
-\frac{1}{4} \Omega_{1,1i}+
\frac{1}{8} (f^2-g^2) \Omega_{1,mn} \varphi_i{}^{mn}=0~.
\label{gatwoxp}
\ey
Furthermore, $24 (\ref{gbonex})+(\ref{gctwox})$ can be written as
\by
&&-5 H_{1mn} \varphi_i{}^{mn}+8 fg F_{mn} \varphi_i{}^{mn}-48 f^2 g^2 \Omega_{i,-+}-24 fg \partial_i (fg)-12 \Omega_{1,1i}\notag\\
&&+ 6 (f^2-g^2) \Omega_{1,mn} \varphi_i{}^{mn}
+(f^2-g^2) H_{kmn}\star\varphi_i{}^{kmn}+ 4 (f^2-g^2) \Omega_{m,n1} \varphi_i{}^{mn}\notag\\
&&- 2 \Omega_{k,mn}\star\varphi_i{}^{kmn}+ 4 \Omega_{k,i}{}^k=0~.
\label{honectwop}
\ey
Similarly,
$(\ref{gbtwox})-\frac{1}{8} (\ref{gconex})$ yields
\by
&&\frac{1}{8} H_{kmn}\star\varphi_i{}^{kmn}+2fg F_{1i}+\frac{1}{4} \partial_i (f^2-g^2)-\frac{1}{8} (f^2-g^2) H_{1mn} \varphi_i{}^{mn}
+\frac{1}{2} (f^2-g^2) \Omega_{1,1i}
\cr &&~~~~
-\frac{1}{4} \Omega_{1,mn} \varphi_i{}^{mn}+\frac{1}{2} \Omega_{m,n1} \varphi_i{}^{mn}-\frac{1}{4} (f^2-g^2)\Omega_{k,mn}\star\varphi_i{}^{kmn}
\cr&&
~~~~~~~~~~~~~+\frac{1}{2}
(f^2-g^2) \Omega_{k,i}{}^k=0~.
\label{hconep}
\ey

The last two equations can be re-arranged as

\by
&&(\frac{1}{10}+\frac{1}{10}f^2 g^2)H_{kmn}\star\varphi_i{}^{kmn}+2fg F_{1i}+\frac{1}{4} \partial_i (f^2-g^2) +\frac{3}{5} (f^2-g^2) fg\partial_i(fg)\cr
&&
+\frac{4}{5} (f^2-g^2) \Omega_{1,1i}+(-\frac{2}{5}+\frac{3}{5} f^2 g^2) \Omega_{1,mn} \varphi_i{}^{mn}+\frac{2}{5}(1+f^2 g^2) \Omega_{m,n1}  \varphi_i{}^{mn}
\cr &&
-\frac{1}{5} (f^2-g^2) \Omega_{k,mn} \star\varphi_i{}^{kmn}+\frac{2}{5} (f^2-g^2) \Omega_{k,i}{}^k
\cr
&&~~~~~ -\frac{1}{5} (f^2-g^2) f g F_{mn}  \varphi_i{}^{mn}+\frac{6}{5} f^2 g^2 (f^2-g^2)\Omega_{i,-+}=0~,
\label{hthrees}
\ey

and
\by
&&-4(1+f^2 g^2) H_{1mn}  \varphi_i{}^{mn}+8 fg F_{mn} \varphi_i{}^{mn}-48 f^2 g^2 \Omega_{i, -+}-8 \partial_i(f^2 g^2)+16 (-1+f^2 g^2) \Omega_{1,1i}
\cr
&&~~~~~
+ 8 (f^2-g^2)
\Omega_{1,mn} \varphi_i{}^{mn}-8 f^2 g^2 \Omega_{k,mn} \star\varphi_i{}^{kmn}
\cr&&~~~~~~~~~~~+16 f^2 g^2 \Omega_{k,i}{}^k-16 (f^2-g^2) fg F_{1i}=0~.
\label{honetwos}
\ey

To continue, eliminate the $G$ fluxes from the eqns above involving the dilaton and also the dilaton from the second equation to find
\by
&&\rd\Phi_i- fg F_{mn} \varphi_i{}^{mn}-\frac{1}{2} (f^2-g^2) H_{-+i}+\frac{1}{2} H_{1mn}  \varphi_i{}^{mn}-\frac{1}{12} (f^2-g^2) H_{kmn} \star\varphi_i{}^{kmn}
\cr &&
~~~~~+
\Omega_{1,1i}-\frac{1}{2} (f^2-g^2) \Omega_{1,mn}  \varphi_i{}^{mn}=0~,
\label{dphic}
\ey
and, after using (\ref{hconep}), to get
\by
&&(f^2-g^2) fg F_{mn}  \varphi_i{}^{mn}- 2 f^2 g^2 H_{-+i}-\frac{1}{2}(f^2-g^2) H_{1mn} \varphi_i{}^{mn}-\frac{1}{3} f^2 g^2 H_{kmn} \star\varphi_i{}^{kmn}
\cr
&&
-2(f^2-g^2) \Omega_{1,1i}+(1-2f^2g^2) \Omega_{1,mn}  \varphi_i{}^{mn} - 2 f g F_{1i} = 0~.
\label{dphid}
\ey



Substituting in (\ref{dphic}) and (\ref{dphid}) for the $H$ fluxes, we find
\by
&&\rd\Phi_i- fg \frac{1+2 f^2 g^2}{6( 1+f^2g^2)} F_{mn}\varphi_i{}^{mn}+\frac{-1+f^2g^2}{3( 1+f^2 g^2)} \Omega_{1,1i}+ \frac{f^2-g^2}{6(1+f^2 g^2)} \Omega_{1, mn}\varphi_i{}^{mn}
\cr&&
-\frac{1}{6( 1+f^2g^2)} \partial_i (f^2g^2)-\frac{1+2 f^2 g^2}{1+f^2 g^2} \Omega_{i,-+}-\frac{1+2 f^2 g^2}{6( 1+f^2g^2)}
\Omega_{k,mn}\star\varphi_i{}^{kmn}
\cr &&
+\frac{1+2 f^2 g^2}{3( 1+f^2g^2)} \Omega_{k,i}{}^k- \frac{(f^2-g^2) fg}{3(1+f^2g^2)} F_{1i}+\frac{1}{3} (f^2-g^2) \Omega_{m,n1} \varphi_i{}^{mn}=0~,
\label{dphicx}
\ey
and
\by
&&\frac{(f^2-g^2) (fg)^3}{3 (1+f^2g^2)} F_{mn} \varphi_i{}^{mn} -\frac{10 (fg)^3}{3(1+f^2g^2)} F_{1i}-\frac{(1-\frac{2}{3} f^2 g^2) f^2 g^2}{2(1+f^2 g^2)}
\partial_i (f^2-g^2)
\cr
&&
+\frac{2 f^2 g^2 (f^2-g^2)}{1+f^2 g^2} \Omega_{i,-+}- \frac{4 f^2 g^2 (f^2-g^2)}{3( 1+f^2 g^2)} \Omega_{1,1i}+\frac{5 f^2 g^2}{3 (1+f^2g^2)} \Omega_{1,mn} \varphi_i{}^{mn}
\cr
&&
+\frac{(f^2-g^2) f^2 g^2}{3 (1+ f^2 g^2)} \Omega_{k,mn} \star\varphi_i{}^{kmn}-\frac{2(f^2-g^2) f^2 g^2}{3 (1+ f^2 g^2)}\Omega_{k,i}{}^k+ \frac{4}{3} f^2 g^2 \Omega_{m,n1} \varphi_i{}^{mn}=0~,~~~
\label{dphidx}
\ey
respectively.
These two last equations can be used to determine the $F_{1i}$ and $F_{mn} \varphi_i{}^{mn}$ fluxes in terms of the dilaton and geometry.  Therefore, they do
not give any additional geometric constraints. Of course on can solve for these $F$ fluxes and the substitute back.   However, the above expressions of the fluxes
are rather economical and we shall not pursue the solution of the linear system further.

\subsection{Solution of the rest of  2-form conditions}
 It remains to investigate the ${\bf 14}$ representation content of the (\ref{Daseq3}) and (\ref{Daseq4}) conditions as the ${\bf 7}$ representation has already been taken into account.
In particular,  one finds that
 \by
  G_{-+ij}^{\bf 14}- (f^2-g^2) F_{ij}^{\bf 14}-4 fg\Omega_{[i,j]1}^{\bf 14}=0~,
  \label{gplusminusf}
 \ey
 and
 \by
  H_{1ij}^{\bf 14}+2 fg F_{ij}^{\bf 14}-2 (f^2-g^2)\Omega_{[i,j]1}^{\bf 14}=0~.
 \label{honef}
 \ey
 These clearly determine components of the $G$ and $H$ fluxes in terms of those of $F$ and the geometry.
 To derive these, we have used that the projections onto the ${\bf 7}$ and ${\bf 14}$ representations are
 \by
 P^{\bf 7}=\frac{1}{3}\big( \delta^{ij}_{mn}-\frac{1}{2} \star\varphi^{ij}{}_{mn}\big)~,~~~P^{\bf 14}=\frac{2}{3}\big( \delta^{ij}_{mn}+\frac{1}{4} \star\varphi^{ij}{}_{mn}\big)~,
\ey
respectively.

\subsection{Solution of the rest of symmetric traceless conditions}

The solution of the conditions (\ref{dst3}) and (\ref{dst4}) gives
\by
\tfrac{1}{6} fg G_{(i }{}^{k_1 k_2 k_3} (\star\varphi)_{j)_o k_1 k_2 k_3} +\tfrac{1}{4}  H_{(i}{}^{k_1 k_2}\varphi_{j)_o k_1 k_2}\qquad\qquad\notag\\
 -  \Omega_{(i,j)_o1} -\tfrac{1}{2} (f^2- g^2) \Omega_{(i,}{}^{k_1 k_2}\varphi_{j)_o k_1 k_2} =0~,
 \label{gsf}
 \ey
 and
 \by
\tfrac{1}{2}fg G_{1 (i}{}^{k_1 k_2} \varphi_{j)_o k_1 k_2} -\tfrac{1}{4} (f^2-g^2) H_{(i}{}^{k_1 k_2}\varphi_{j)_o k_1 k_2}\qquad\qquad\notag\\
 + (f^2-g^2) \Omega_{(i,j)_o1} +\tfrac{1}{2}  \Omega_{(i,}{}^{k_1 k_2}\varphi_{j)_o k_1 k_2} =0~.
 \label{gst}
\ey
This concludes the solution of all conditions which arise from the KSEs.

\section{Special case \texorpdfstring{$f=g$}{f = g}}
\label{appspec}

\subsection{The + components of the fluxes and geometry}

The conditions on the geometry can be read off from  those in  (\ref{plusgeom}) after setting $f=g=1/\sqrt{2}$ as
\by
&\Omega_{+,+1}=\Omega_{+,-+}=\Omega_{1,1+}=0 ~,\\
&\Omega_{+, 1 i} = \Omega_{1,+ i} = \Omega_{i,+1} =\Omega_{+,+i}=0~,
\\
&\Omega_{+,}{}^{k_1 k_2} \varphi_{i k_1 k_2}  = 0 ~,~~~
  \Omega_{k_1,k_2 +} \varphi^{k_1 k_2}{}_i = 0 ~, ~~~
  \Omega_{(i,j)+}=0~,
\ey
and those for the fluxes can be derived from  (\ref{splus}), (\ref{vplus}) and (\ref{restplus}) as
\by
&G_{+ijk}=0~,~~~F_{+i}=0~,~~~H_{+ij}=0~,~~~H_{+1i}=0~,
\\
&F_{+1}=0~,~~~G_{+1ij}\varphi^{ij}{}_k=0~,
\\
&\rd\Phi_+=\rd f_+=\rd g_+=0~,
\\
&G_{+1ij}=2 \Omega_{i,j+}~.
\ey
Observe that many more components vanish compared to the generic case.

\subsection{Solution of the rest of  scalar conditions}

The scalar conditions imply
\by
G_{1ijk} \varphi^{ijk}=H_{-+1}=F_{-+}=0~,~~~ \Omega_{-,- +} =\Omega_{-,+1}+\Omega_{1,+-}=\Omega_{i,jk} \varphi^{ijk}=0~,
\ey
and
\by
&&d\Phi_1+\frac{1}{3} \Omega_{k,1}{}^k-\frac{4}{3} \Omega_{-,+1}+\frac{1}{3} S=0~,~~~\frac{1}{24} G_{ijkl} \star\varphi^{ijkl} + 4 \Omega_{-,+1}+S=0~,
\\
&&
-\frac{1}{2} H_{ijk}\varphi^{ijk}- 2 \Omega_{k,1}{}^k+ 2 \Omega_{-,+1}+ 4 S=0~,~~ F_{-1} - \frac{1}{6} G_{- i j k } \varphi^{i j k} = 0~.
\label{dddphi}
\ey
This is a significant simplification compared to the solution in the generic case.

\subsection{Solution of the rest of  vector conditions}

In particular, the solution is
\by
&&\Omega_{i,+-}+\Omega_{-,+i}=0~,~~~H_{-+i}=0~,
\ey
and
\by
&&\frac{1}{6} G_{imnl} \varphi^{mnl}+ \Omega_{1,mn} \varphi_i{}^{mn}=0~,~~~-\frac{1}{2} G_{-+mn}\varphi_i{}^{mn}+F_{1i}- \Omega_{1,mn} \varphi_i{}^{mn}=0~,
\cr
&&H_{mnl} \star\varphi_i{}^{mnl}+ 8 F_{1i}+ 4 \Omega_{m,n1} \varphi_i{}^{mn}- 2\Omega_{1,mn} \varphi_i{}^{mn}=0~,
\cr
&&
-5 H_{1mn}\varphi_i{}^{mn}-2\Omega_{m,nl} \star\varphi_i{}^{mnl}+ 4 F_{mn} \varphi_i{}^{mn}-12 \Omega_{i, -+}-12 \Omega_{1,1i} +4 \Omega_{k,i}{}^k=0~,
\cr
&&
-G_{-+1i}-\frac{4}{5} \Omega_{i, -+}+\frac{1}{10}   F_{mn} \varphi_i{}^{mn}-\frac{4}{5} \Omega_{1,1i}+\frac{1}{5}   \Omega_{m,nl} \star\varphi_i{}^{mnl}-\frac{2}{5}\Omega_{k,i}{}^k=0~,
\cr
&&
-\frac{1}{24} G_{1mnl} \star\varphi_i{}^{mnl}-\frac{1}{10}   F_{mn} \varphi_i{}^{mn}+\frac{4}{5} \Omega_{i, -+}-\frac{1}{5} \Omega_{1,1i}+\frac{1}{20}\Omega_{m,nl} \star\varphi_i{}^{mnl}
-\frac{1}{10}\Omega_{k,i}{}^k=0~,
\cr
&&
\rd\Phi_i-\frac{1}{10} F_{mn} \varphi_i{}^{mn}-\frac{6}{5} \Omega_{i, -+}-\frac{1}{5} \Omega_{m,nl} \star\varphi_i{}^{mnl}-\frac{1}{5} \Omega_{1,1i}+\frac{2}{5}\Omega_{k,i}{}^k=0~,
\cr
&&
-F_{1i}+ \Omega_{m,n1} \varphi_i{}^{mn}+ \Omega_{1,mn} \varphi_i{}^{mn}=0~.
\ey
Furthermore, we have
\by
&&\frac{1}{2} G_{-1mn} \varphi_i{}^{mn} +F_{-i}-H_{-1i}-\Omega_{-,mn} \varphi_i{}^{mn}=0~,
\cr
&&-\frac{1}{6} G_{-mnl} \star\varphi_i{}^{mnl}-\frac{1}{2} H_{-mn}\varphi_i{}^{mn}-2 \Omega_{-,1i}=0~.
\ey
In fact, some of the above equations can be simplified further to yield
\by
&&-\frac{1}{2} G_{-+mn}\varphi_i{}^{mn}+ \Omega_{m,n1} \varphi_i{}^{mn}=0~,~~~
\cr
&&
H_{mnl} \star\varphi_i{}^{mnl}+ 12  \Omega_{m,n1} \varphi_i{}^{mn}+6\Omega_{1,mn} \varphi_i{}^{mn}=0~.
\ey
Observe that we have been able to solve completely the linear system for the vector fluxes.

\subsection{Solution of the rest of  2-form and  symmetric traceless conditions}

 It remains to investigate the ${\bf 14}$ content of  (\ref{Daseq3}) and (\ref{Daseq4}) as the ${\bf 7}$ representation has been taken into account.
 One finds that
 \by
  G_{-+ij}^{\bf 14}-2\Omega_{[i,j]1}^{\bf 14}=0~,~~~
  H_{1ij}^{\bf 14}+ F_{ij}^{\bf 14}=0~.
 \ey
Similarly, the conditions (\ref{dst3}) and (\ref{dst4}) give
\by
&&\tfrac{1}{12}  G_{(i }{}^{k_1 k_2 k_3} \star\varphi_{j)_o k_1 k_2 k_3} +\tfrac{1}{4}  H_{(i}{}^{k_1 k_2}\varphi_{j)_o k_1 k_2}-  \Omega_{(i,j)_o1}  =0~,\cr
&&\tfrac{1}{2}G_{1 (i}{}^{k_1 k_2} \varphi_{j)_o k_1 k_2} +  \Omega_{(i,}{}^{k_1 k_2}\varphi_{j)_o k_1 k_2} =0~.
\ey
This concludes the solution of the linear system for the special case.

\end{document}